\documentclass{article}

\usepackage{xspace}

\usepackage{arxiv}

\usepackage[utf8]{inputenc} 
\usepackage[T1]{fontenc}    
\usepackage{hyperref}       
\usepackage{url}            
\usepackage{booktabs}       
\usepackage{amsfonts}       
\usepackage{nicefrac}       
\usepackage{microtype}      
\usepackage{lipsum}
\usepackage{graphicx}
\graphicspath{ {./images/} }

\usepackage{amsmath}
\usepackage{multirow}
\usepackage{graphicx}
\usepackage{adjustbox}
\usepackage{pifont}
\usepackage{svg}
\usepackage{float}   
\usepackage{xspace}
\usepackage{wrapfig}
\usepackage{natbib}
 
\newcommand{\model}{\textsc{Nef-Net+}\xspace}
\newcommand{\dataset}{{Panobench}\xspace}
\newcommand{\pretrain}{{Any-Pairs Pretraining}\xspace}
\newcommand{\hospital}{{Device Calibration}\xspace}
\newcommand{\patient}{{On-the-fly Calibration}\xspace}

\newcommand{\eg}{\emph{e.g.},\xspace}

\title{\model: Adapting Electrocardio Panorama \\ in the wild}

\author
{
Zehui Zhan$^{1,2,*}$
\and
Yaojun Hu$^{3*}$\and
Jiajing Zhan$^{4}$\and
Wanchen Lian$^{1}$\and
Wanqing Wu$^{2\dagger}$\and
Jintai Chen$^{1\dagger}$\\
$^1$Information Hub, Hong Kong University of Science and Technology (Guangzhou), Guangzhou, 511458, China\\
$^2$School of Biomedical Engineering, Sun Yat-sen University, Shenzhen, 518107, China
\\
$^3$College of Computer Science and Technology, Zhejiang University, Hangzhou, 310012, China\\
$^4$Department of Electrical and Electronic Engineering, University of Hong Kong, Pokfulam, Hong Kong\\
jintaiCHEN@hkust-gz.edu.cn
}
\begin{document}
\maketitle
\begin{abstract}
Conventional multi-lead electrocardiogram (ECG) systems capture cardiac signals from a fixed set of anatomical viewpoints defined by lead placement. 
However, certain cardiac conditions (\eg Brugada syndrome) require additional, non-standard viewpoints to reveal diagnostically critical patterns that may be absent in standard leads. 
To systematically overcome this limitation, Nef-Net was recently introduced to reconstruct a continuous electrocardiac field, enabling virtual observation of ECG signals from arbitrary views (termed \textit{Electrocardio Panorama}). 
Despite its promise, Nef-Net operates under idealized assumptions and faces in-the-wild challenges, such as long-duration ECG modeling, robustness to device-specific signal artifacts, and suboptimal lead placement calibration.
This paper presents \model, an enhanced framework for realistic panoramic ECG synthesis that supports arbitrary-length signal synthesis from any desired 
view, generalizes across ECG devices, and compensates for operator-induced deviations in electrode placement. 
These capabilities are enabled by a newly designed model architecture that performs direct view transformation, incorporating a workflow comprising offline pretraining, device calibration tuning steps as well as an on-the-fly calibration step for patient-specific adaptation. 
To rigorously evaluate panoramic ECG synthesis, we construct a new \textit{Electrocardio Panorama} benchmark, called \dataset, comprising 5367 recordings with \textbf{48}-view per subject, capturing the full spatial variability of cardiac electrical activity. 
Experimental results show that \model delivers substantial improvements over Nef-Net, yielding an increase of around 6 dB in PSNR in real-world setting. The code and \dataset will be released in a subsequent publication.
\end{abstract}

\section{Introduction}
Cardiovascular diseases remain the leading cause of morbidity and mortality worldwide~\citep{first-killer}, claiming tens of millions of lives each year and imposing profound disability burdens that underscore an urgent clinical imperative~\citep{2023heart-disease-statistics}.
Among diagnostic modalities, electrocardiogram (ECG) has established itself as indispensable, providing a non-invasive, cost-effective approach that offers immediate insights into the complex dynamics of cardiac electrical activity~\citep{value-of-12lead-electrocardiogram}.

The number of ECG observation viewpoints directly correlates with both practical complexity and the comprehensiveness of cardiac condition understanding~\citep{kligfield2007recommendationsofecg}. 
The standard 12-lead ECG, which is widely used in clinical practice, is generally considered a practical compromise between acquisition cost and clinical utility~\citep{holter1961new}. The electrode placement for the standard 12-lead ECG is illustrated in Figure~\ref{fig:position}(a).
However, this conventional setup is still insufficient for detecting certain cardiac pathologies with specialized localization patterns, and may require additional, non-standard viewpoints. For instance, posterior myocardial infarction often requires additional posterior leads (V7-V9)\footnote{\scriptsize V7–V9 are ECG leads, beyond the standard 12-lead system, placed on the back to record activity from the posterior heart wall.} for definitive diagnosis~\citep{van2007posteriormyocardial}.
Brugada syndrome detection requires ECG acquisition from additional viewpoints (at 2nd/3rd ICS)\footnote{\scriptsize The 2nd/3rd ICS are specific anatomical locations on the chest.}, as these viewpoints uniquely capture the pathological signals caused by right ventricular outflow tract depolarization abnormalities~\citep{berne2012brugada}. 

To address the trade-off between information richness and view availability in ECGs and to advance the clinical utility of panoramic ECG analysis, \cite{2021-nefnet-electrocardio-panorama} introduced \textit{Electrocardio Panorama}, which constructs an implicit neural representation, enabling the generation of ECG signals from arbitrary viewpoints in real time. While this pioneering approach lays an important foundation, Nef-Net nonetheless shows several limitations that constrain its adoption:

\begin{table}[t] 
    \centering
    \caption{Comparison of Nef-Net \& \model. Synthesis performance tested on in-the-wild ECGs. The synthesis performance of the original Nef-Net model is re-evaluated under the same patient-level testing protocol as Nef-Net+, instead of the original heartbeat-level in original paper.}
    \begin{tabular}{ccccccc}
\toprule
Method  & ECG Length & Device & Lead Placement & Synthesis: PSNR($\uparrow$) \\
\midrule
Nef-Net& Heartbeat & Restricted & High-Precision & 24.10-28.01\\
\model& Continuous & Agnostic & In-the-wild & 32.07\textcolor{red}{(7.97$\uparrow$)}-34.82\textcolor{red}{(6.81$\uparrow$)} \\
\bottomrule
    \label{Tabl:COMPARISON}
    \end{tabular}
    \vspace{-0.8cm}
\end{table}

\textbf{(1) Heartbeat-level modeling.}
Nef-Net is confined to single-heartbeat ECG reconstruction, limiting clinical utility since continuous monitoring requires long-duration signals to capture inter-beat dynamics and arrhythmia patterns.
\textbf{(2) Underuse of view-specific features.}
Its encoder-decoder design compresses ECG features to reconstruct cardiac fields but neglects the varying relevance of each view to the query. By uniformly averaging features from all recorded views, the model mixes query-irrelevant information into the representation, leading to blurred and coarse reconstructions. This degradation is further exacerbated when only few views are available for supervision.
\textbf{(3) Neglect of practical deployment factors.}
Training assumes idealized data, and Nef-Net overlooks two real-world challenges: inter-device shifts (\eg variations in sensor characteristics and signal-processing pipelines across ECG devices) and inter-subject differences (\eg electrode placement offsets introduced by clinical staff).
\textbf{(4) Limited validation of panoramic ECG.}
Due to the unavailability of dense-view datasets, evaluation has been restricted to 12-lead ECGs under narrow angular settings, which is insufficient to assess the model’s generalizability in reconstructing the global cardiac field. 
\textit{We make the following contributions to address these limitations:}
\begin{enumerate}
    \item[\textbf{(A)}] \textbf{Geometric View Transformer (GeoVT).} We introduce a geometry-aware cross-attention architecture that explicitly models spatial relationships between query and recorded ECG views, selectively extracting the most relevant features for direct query-view transformation. This enables end-to-end synthesis of arbitrary-length ECG signals, allowing \model to achieve superior performance with fewer parameters than Nef-Net.
    \item[\textbf{(B)}] \textbf{A New Paradigm for Model Development.} We propose a unified three-stage pipeline for developing and deploying \model in real-world Electrocardio Panorama applications. In the initial \textbf{\textit{\pretrain}} stage, the model acquires robust cross-view transformations under controlled laboratory conditions. Then, \textbf{\textit{\hospital}} stage addresses feature distribution shifts arising from heterogeneity across ECG devices in clinical environments. Finally, the \textbf{\textit{\patient}} stage enables rapid, geometry-aware adaptation at the level of individual examinations, compensating for variations in electrode placement.
    \item[\textbf{(C)}] \textbf{A Novel Panoramic ECG Dataset and Superior Performance.} To enable comprehensive benchmarking of Electrocardio Panorama synthesis, we curated \textbf{\textit{\dataset}} (lead positions shown in Figure~\ref{fig:position}(c)), the first dense \textbf{48}-lead ECG dataset with precisely CT-measured spherical coordinates for each view, with inter-subject angular variance quantified across \textbf{5,367} recordings. Experimental results show that \model accurately synthesizes novel ECG views, offering a promising avenue toward more comprehensive clinical ECG assessment.
    
    \begin{figure*}[t]
    \centering  
    \vskip -0 in
    \includegraphics[width=0.9\textwidth]{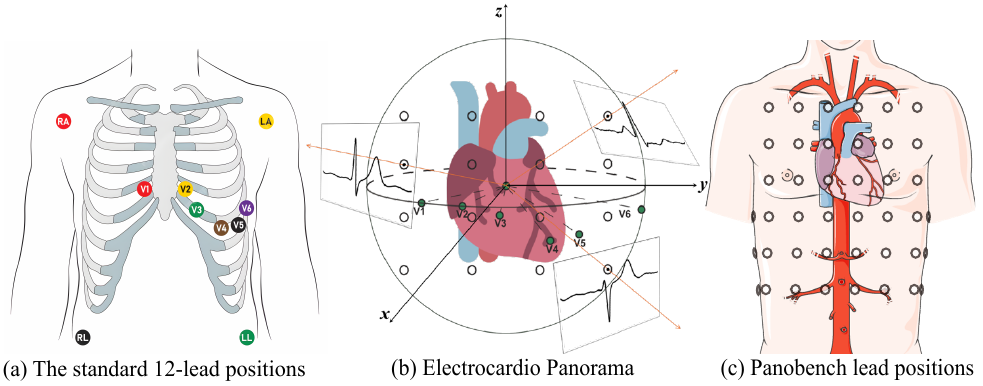}
    \vskip -0 in
    \caption{
(a) Standard 12-lead electrode placement for conventional ECG recording.
(b) \textit{Electrocardio Panorama} enables any user-desired virtual ECG viewpoints for comprehensive visualization of cardiac electrical activity.
(c) The \dataset benchmark, encompassing \textbf{48} distinct ECG viewpoints for each case, enables rigorous evaluation of Electrocardio Panorama generation models.}
\vspace{-10pt}
\label{fig:position}
\end{figure*}
\end{enumerate}

\section{Background and related work}
\subsection{Electrocardiogram (ECG)}
ECG recordings are time-series signals of cardiac electrical activity. Each cardiac cycle can be decomposed into six non-overlapping deflections: the P wave, PR segment, QRS complex, ST segment, T wave, and TP segment. The standard 12-lead ECG protocol (lead positions shown in Figure~\ref{fig:position}(a)) is widely used for cardiovascular screening and typically captures 10-second recordings from 6 limb leads (\uppercase\expandafter{\romannumeral1}, \uppercase\expandafter{\romannumeral2}, \uppercase\expandafter{\romannumeral3}, aVR, aVL and aVF) and 6 chest leads (V1-V6)~\citep{kligfield2007recommendationsofecg}. A more detailed introduction to ECGs is provided in Appendix~\ref{sec:Theoretical support}.



\subsection{ECG view reconstruction and synthesis}
ECG view reconstruction is essential for recovering missing leads and enabling comprehensive cardiac assessment. Early methods relied on linear transformations~\citep{2004-linear}, assuming predominantly linear relationships across leads, which fails to capture the inherently nonlinear cardiac dynamics~\citep{mcculloch1998computationalbiologyoftheheart}.
To address this limitation, nonlinear approaches have been developed, including recurrent neural networks (RNNs)~\citep{2020-rnn}, long short-term memory networks (LSTMs)~\citep{2022-lstm}, convolutional neural networks (CNNs)~\citep{2022-cnn,chen2024multi}, and conditional generative adversarial networks (CGANs)~\citep{seo2022multipleganto12,golany2019pgans}, which better model complex inter-lead relationships. Yet, these methods are limited to reconstructing predefined, known views and cannot generate novel views that may be clinically valuable.
\cite{2021-nefnet-electrocardio-panorama} first introduced the concept of \textit{Electrocardio Panorama}, which enables the synthesis of any unseen views conditioned on viewing angles. While this represents a significant conceptual advance, the approach remains unsuitable for real-world clinical applications due to its neglect of real-world challenges like operational offsets and device inconsistencies. Our method addresses these constraints and provides a more robust solution for panoramic ECG observation.

\section{Methodology}
\subsection{Architecture}
\begin{figure}[t]
    \centering    \includegraphics[width=0.95\linewidth]{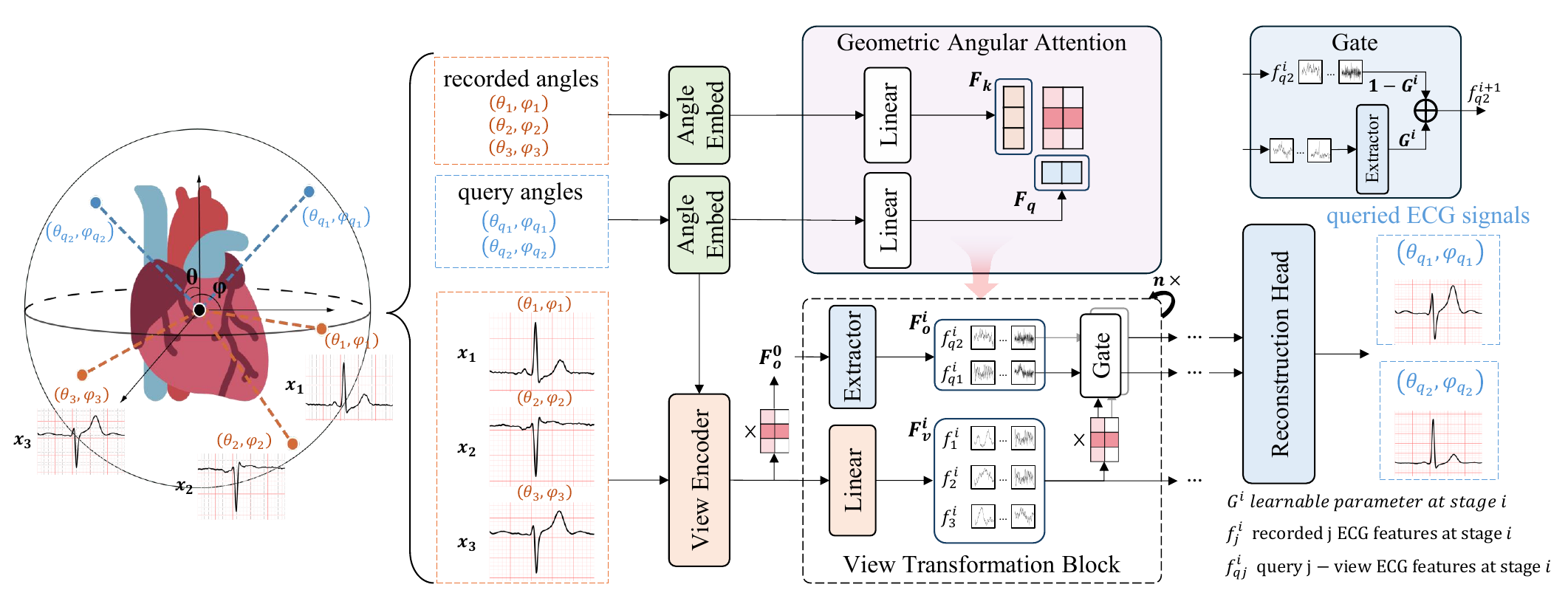}
    \vskip -0 in
    \caption{\textbf{Our proposed \model architecture for Electrocardio Panorama synthesis (illustrated for a 3-input to 2-query view synthesis task as example).} The \model first employs a View Encoder to extract features from the Recorded ECG that are relevant to the Queried ECG. These extracted features are then fused using a Geometric View Transformer to synthesize the query view.}
    \label{fig:archi}
    \vspace{0em}
\end{figure}

The key idea of \model is to formulate ECG view synthesis as a direct view-to-view transformation problem, bypassing neural electrocardio field reconstruction~\citep{2021-nefnet-electrocardio-panorama} and instead exploiting inter-lead spatial dependencies defined by angular relationships. \model incorporates three core components: \textbf{Angle Embedding}, \textbf{View Encoder}, and \textbf{Geometric View Transformer (GeoVT)}, as illustrated in Fig.~\ref{fig:archi}.
Formally, let $X=\{x_1,\cdots, x_l\}$ with each $x_i \in \mathbb{R}^{1\times t}$ denote $l$ ECG signals recorded from distinct viewing angles. 

\paragraph{Angle Embedding.} We simply extend the ``Angular Encoding'' module of Nef-Net with an additional linear projection to align feature dimensions, mapping recorded angles $A_k$ and query angles $A_q$ into a higher-dimensional angle space, respectively.


\paragraph{View Encoder.}
Each recorded single-lead ECG signal $x_i \in \mathbb{R}^{1\times t}$ is processed by a 1-D ResNet basic block following~\citep{2021-nefnet-electrocardio-panorama}, yielding $f_{x_i}\in\mathbb{R}^{c\times t'}$. This representation is then concatenated with the query feature $F_q$ in a FiLM-style affine modulation~\citep{perez2018film}, which amplifies signal features aligned with $(\theta_q,\varphi_q)$ and suppresses irrelevant ones. Aggregating the outputs from all $l$ recorded leads produces the encoded feature matrix $F_v^0=[f_1,\ldots,f_l]$.

\paragraph{Geometric View Transformer (GeoVT).} As each ECG view conveys partially redundant yet complementary information about the query view~\citep{pipberger1961correlationofdifferentleads}, our GeoVT is designed with three key components: a Geometric Angular Attention Module ($\texttt{M}_{\text{GAA}}$), a View Transformation block, and a Reconstruction head. $\texttt{M}_{\text{GAA}}$ estimates geometric similarity between recorded and query views, the View Transformation block projects recorded features into query-aligned representations, and the Reconstruction head decodes them to reconstruct target-view ECG signals.

(I) \textit{Geometric Angular Attention Module ($\texttt{M}_{\text{GAA}}$)}. The $\texttt{M}_{\text{GAA}}$ implements a cross-attention mechanism~\citep{crossAttention} that compares the angular embedding of the query leads $F_q$ with recorded leads $F_k$. Formally, the Geometric Angular Attention map (GAA) is computed by:
\begin{equation}
    GAA = \text{softmax}\left(\frac{F_q W_q (F_k W_k)^{\top}}{\sqrt{d'}}\right)
    \label{eq1}
\end{equation}
where $W_q, W_k \in \mathbb{R}^{d\times d'}$ are learnable projection matrices.

(II) \textit{View Transformation Block.} In GeoVT, we stack $L$ view transformation blocks to transfer hierarchical features from recorded signals to the query signals. In block $i$, the recorded signal features $F_v^i$ are projected into an angular latent space by $F_v^{i+1} = \text{Linear}(F_v^i)$ and fused according to the GAA. The resulting representations are integrated block by block through a spatial gating mechanism, by: 
\begin{equation}
    F_o^{i+1} = F_o^i\odot(1-G^i) + \textit{Ext.}(F_v^{i}\times GAA)\odot G^i 
\end{equation}
where $G^i$ is a learnable parameter with a sigmoid function. The feature extractor (\textit{Ext.}) follows the design of SE blocks~\citep{hu2018squeeze}.
Through this hierarchical process, GeoVT progressively refines features from the recorded ECG signals in a coarse-to-fine manner, focusing on those relevant to the query view and enabling effective cross-view transformation. Notably, all blocks share the same GAA map as defined in Eq.~\ref{eq1}.

(III) \textit{reconstruction head.} 
The reconstruction head maps the fused embeddings $F_o^L$ back to the time domain using a sequence of upsampling blocks. Each block performs linear interpolation, followed by a convolution module that incorporates a spectral-normalized 1D convolution~\citep{miyato2018spectral}, layer normalization, and a GELU activation.

\subsection{\model Development and Deployment}
\begin{figure}[t]
    \centering
    \includegraphics[width=0.95\linewidth]{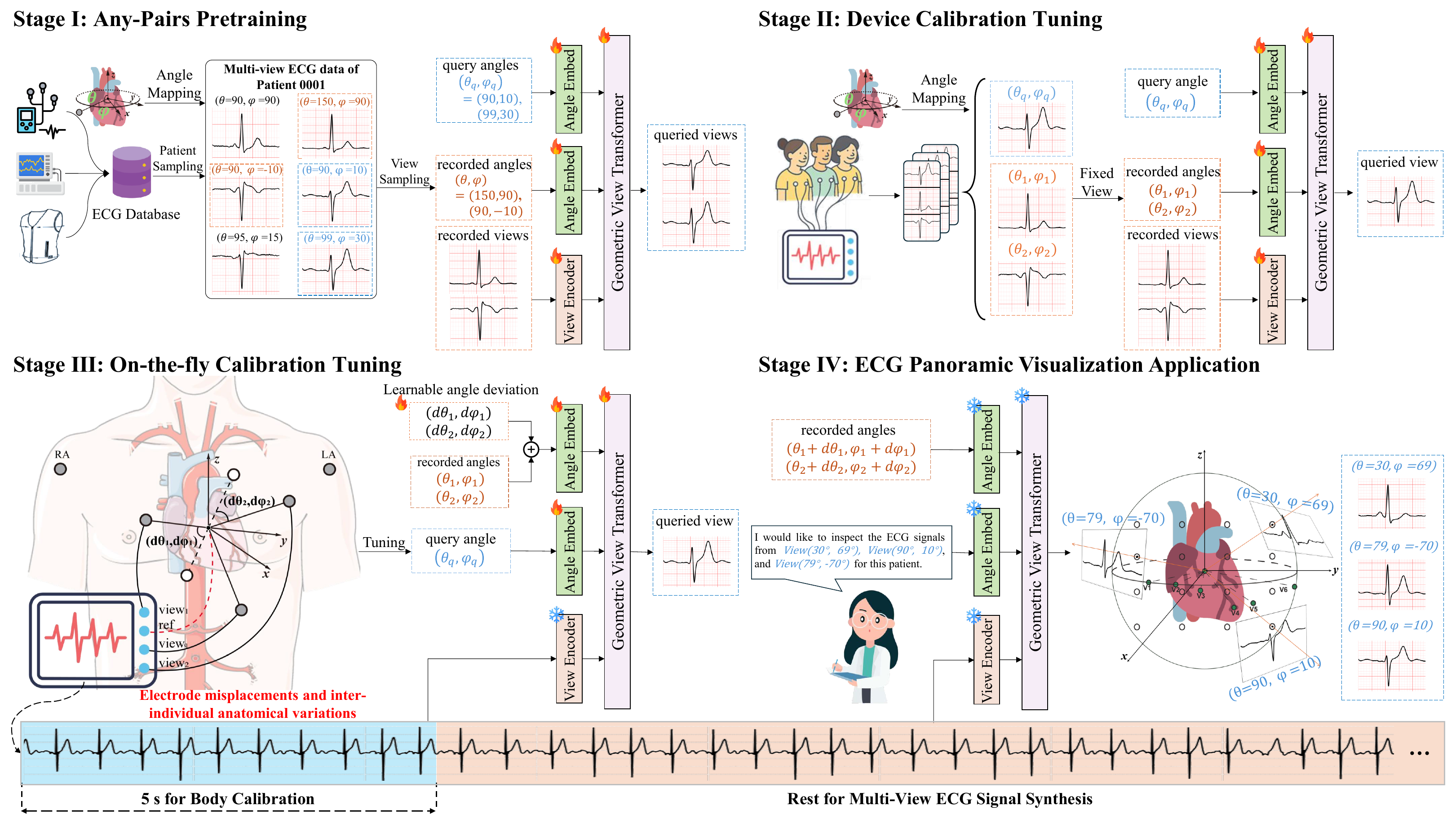}
    \vskip -0.1 in
    \caption{The Development (Stage I, II, III) and Deployment (Stage IV) Workflow of \model.}
    \label{fig:learning}
    \vspace{-0.5em}
\end{figure}
To deploy an ECG analysis model in real-world settings, a central challenge lies in the heterogeneity of recording devices~\citep{bailey1990recommendationsstandardizationandspecificationsinautomatedelectrocardiography} and the variability arising from operator-dependent procedures and patient-specific variations~\citep{schijvenaars2008intraindividualdifference}. The original Nef-Net overlooks device-specific discrepancies and case-specific variability. To ensure that our \model remains robust and clinically applicable, we design a three-stage development and deployment strategy: (1) a device- and case-agnostic pretraining stage, called \pretrain (\textsc{AnyPre}), which lets \model to learn invariant spatiotemporal representations across views under laboratory conditions; (2) \hospital (\textsc{D-Cal}), a device-specific calibration stage that adapts \model to different ECG devices during deployment; and (3) \patient (\textsc{OF-Cal}), a case-adaptive calibration stage applied at each examination to align the model with case-level variability. 
\paragraph{Stage I: \pretrain}
In this phase, \model is pretrained on ECG cases collected from heterogeneous devices to learn fundamental ECG patterns. For each case, the available ECG views are randomly partitioned into two subsets to form the recorded–query pairs $(X_i, Y_i)$. \textit{All} parameters are kept trainable, and we dynamically sample these recorded–query pairs in model training. Notably, in accordance with ECG principles, two limb leads are consistently designated as recorded signals, as they provide essential reference potentials for constructing the remaining leads. Following~\citep{2021-nefnet-electrocardio-panorama}, we optimize the network with the mean absolute error (MAE) loss, defined as $\mathcal{L}_{\text{MAE}}= \parallel \hat{Y_i} - Y_i \parallel_1$, where $\hat{Y_i}$ are synthesized by the model.

This strategy exposes the model to diverse sensor characteristics and patient demographics, mitigating protocol-specific overfitting while compelling it to infer cardiac dynamics from arbitrary lead combinations, thereby enhancing robustness and generalization to unseen electrode configurations.

\paragraph{Stage II: \hospital} 
Due to variations in ECG acquisition protocols across devices—including hardware design, electrode materials, and others—we present a \hospital stage for local adaptation. In this stage, the model is fine-tuned on all recorded–query pairs from the target device using $\mathcal{L}_{\text{MAE}}$, markedly improving alignment with the specific hardware configuration.

\paragraph{Stage III: \patient} 
Although an ideal viewing angle is defined by~\citep{2021-nefnet-electrocardio-panorama} and this paper, the actual recorded ECG angles $(\theta_{\mathrm{real}}, \varphi_{\mathrm{real}})$ often deviate substantially from the ideal $(\theta, \varphi)$ due to two primary factors: (1) \textbf{electrode placement variability} arising across ECG examinations, causing viewing angle deviations; and (2) \textbf{inter-subject anatomical variability}, such as differences in heart position, which introduce subject-specific angular offsets from the population mean for each view (\eg a standard deviation of up to $10.6^\circ$ can be observed in \dataset). \textbf{\textit{However, the precise offsets are generally difficult to obtain directly.}} To compensate for these discrepancies, we introduce learnable angular deviation parameters $(d\theta, d\varphi)$, which are added to the ideal angles to form $(\theta + d\theta, \varphi + d\varphi)$, enabling \model to dynamically adjust for both sources of variation. In accordance with clinical ECG recording standards (minimum 10-second duration), the initial 5-second segment is allocated for model calibration on the fly. 
During this stage, the \textit{View Encoder} and \textit{Reconstruction Head} parameters remain frozen, while fine-tuning adapts the angle embeddings to individual-specific deviations.

\subsection{\dataset: A dense benchmark for \textit{Electrocardio Panorama}}
To enable comprehensive evaluation of panoramic ECG view synthesis, we curated a new benchmark, \textbf{\textit{\dataset}}, which for the first time extends beyond previous datasets limited to only \textbf{8} or \textbf{12} views. \dataset comprises \textbf{9,360} ten-second recordings with \textbf{48} viewpoints (\textbf{6} limb and 42 precordial leads), each annotated with CT-derived spherical coordinates $(\theta,\varphi)$ following~\citep{2021-nefnet-electrocardio-panorama}. 
By expanding to 48 leads (views), \dataset provides a high-resolution representation of cardiac electrical dynamics, enabling signal analysis from diverse perspectives. This design supports rigorous validation of panoramic ECG synthesis methods and establishes a foundation for clinical translation. In contrast, traditional \textbf{12}-lead settings require input, supervision, and synthesis to be partitioned among the same limited leads, making validation inherently less comprehensive.


\section{Experiments}
\subsection{Datasets and Implementation Details}
We conduct experiments on the PTBXL dataset~\citep{wagner2020ptb}, Tianchi ECG dataset\footnote{https://tianchi.aliyun.com/competition/entrance/231754/information?lang=en-us}, CPSC2018 dataset~\citep{cpsc2018}, ChinaDB dataset~\citep{chinadb}, and our curated dataset-\dataset.
Each dataset is randomly split into 80 percent training and 20 percent testing. Detailed information about the datasets is provided in Appendix~\ref{sec:Detailed Information of ECG Dataset}. 

In the \textit{\pretrain} stage, the model is trained on the combined training sets of four public ECG datasets, with evaluation conducted on their respective test sets.
In the \textit{\hospital} stage, a single dataset with fixed input-lead configurations is utilized for both training and testing.
In the \textit{\patient} stage, the first 5-second segment of each patient's ECG recording is used for model adaptation, with the subsequent 5-second segment reserved for performance evaluation. 

All experiments are implemented using PyTorch 1.9 on three NVIDIA RTX2080Ti GPUs, each with 11 GB of memory.
The implementation details of each experiment are shown in Appendix~\ref{sec:Implementation Details of Different Experiments}.

\subsection{Performance Evaluation}
\label{comparison with related works}
In this section, we evaluate \model's ability to generate an \textit{Electrocardio Panorama} via two complementary tasks: \textit{reconstruction}, in which the model regenerates ECG signals from viewpoints used in training, and \textit{synthesis}, in which it produces signals for entirely unseen viewpoints, following~\citep{2021-nefnet-electrocardio-panorama}. Signal quality is assessed using peak signal-to-noise ratio (PSNR) and structural similarity index (SSIM). For the \textit{reconstruction} task, Nef-Net~\citep{2021-nefnet-electrocardio-panorama}, E-LSTM~\citep{E-LSTM}, and KIM~\citep{KIM} were evaluated following the settings of previous work. 
\textit{Owing to the use of longer ECG recordings rather than beat-level data, our Nef-Net reproduction shows degraded performance compared to the original.}
For the \textit{synthesis} task, we compare only with Nef-Net, as it is the only existing method capable of generating signals for unseen viewpoints. The number of views for input, supervision (reconstruction), and generation are listed in parentheses, in that order.

\begin{table}[H]
\vspace{-0.5em}
  \centering
  \caption{Performance on \textit{reconstruction} and \textit{synthesis} tasks on the Tianchi, Chinadb, CPSC2018 and PTBXL datasets. In the synthesis tasks, the numbers of views for input, reconstruction, and synthesis are orderly listed in parentheses. The
best performances are highlighted in \textbf{bold}.}
\resizebox{0.9\columnwidth}{!}{
    \begin{tabular}{lc|cc|cc|cc|cc}
\toprule
& \multirow{2}[4]{*}{} & \multicolumn{2}{c|}{ChinaDB} & \multicolumn{2}{c|}{CPSC2018} & \multicolumn{2}{c|}{Tianchi} & \multicolumn{2}{c}{PTBXL} \\
\cmidrule(l){3-10}
&& PSNR  & SSIM  & PSNR  & SSIM  & PSNR  & SSIM  & PSNR  & SSIM \\
\midrule
\multicolumn{10}{c}{\rule{0pt}{0em} \textit{View Reconstruction}}\\ 
\midrule
E-LSTM &(3,9) & 20.56 & 0.811  & 21.37 & 0.824 & 22.76 & 0.848 & 20.04 & 0.810 \\
Nef-Net &(3,9) & 29.59 & 0.961  & 29.12 & 0.958 & 31.44 & 0.965 & 30.22 & 0.962 \\
\model &(3,9)&\textbf{35.84}&\textbf{0.977}&\textbf{36.12}&\textbf{0.981}&\textbf{37.13}&\textbf{0.982}&\textbf{35.21}&\textbf{0.974}\\
\midrule
KIM &(8,12) & 27.65 & 0.952 & 27.82 & 0.956 & 28.01 & 0.937 & 26.71 & 0.929 \\
Nef-Net &(8,12) & 32.74 & 0.967  & 31.68 & 0.971 & 33.72 & 0.961 & 30.58 & 0.977  \\
\model &(8,12) &\textbf{39.54}&\textbf{0.978}&\textbf{38.69}&\textbf{0.981}&\textbf{41.52}&\textbf{0.983}&\textbf{39.15}&\textbf{0.983}\\
\midrule
\multicolumn{10}{c}{\rule{0pt}{0em}\textit{Unseen View Synthesis}}\\ 
\midrule
Nef-Net& (3,8,1)& 25.24 & 0.951  & 26.72 & 0.957 & 27.92 & 0.959 & 24.10 & 0.922 \\
\model& (3,8,1)&\textbf{32.57}&\textbf{0.981}&\textbf{33.62}&\textbf{0.985}&\textbf{34.46}&\textbf{0.976}&\textbf{33.41}&\textbf{0.982}\\
\midrule
Nef-Net& (5,6,1)& 26.06 & 0.954  & 26.11 & 0.948 & 28.01 & 0.959 & 25.37 & 0.942\\
\model&  (5,6,1)& \textbf{33.16} &\textbf{0.982} &\textbf{32.76} &\textbf{0.982}&\textbf{34.82}&\textbf{0.977}&\textbf{32.07}&\textbf{0.986}\\
\bottomrule
    \end{tabular}%
}
  \label{tab:comparison_dataset}
  \vspace{-0.5em}
\end{table}

As shown in Table~\ref{tab:comparison_dataset}, for \textit{reconstruction}, our \model significantly outperforms previous methods, and these results can serve as a reference for the model’s \textit{synthesis} capability. The lower portion of the table shows that \model's \textit{synthesis} performance nearly matches its reconstruction performance, highlighting its robust ability to generalize beyond observed view distributions, which is critical for clinical applicability where novel configurations are common. Moreover, \model consistently outperforms Nef-Net across all datasets, yielding substantially higher PSNR and SSIM in \textit{Electrocardio Panorama} synthesis. A detailed analysis of the factors contributing to these gains is provided in Section~\ref{electrocardio panorama}.

\begin{table}[t]
\centering
\vskip -0.3 in
\caption{Performance Comparison of Nef-Net and \model on ECG Panoramic Synthesis Across Varying Numbers of Input Leads and Supervised Leads on the \dataset dataset. In the table configuration, leads I and II belong to the standard bipolar limb lead system, which measures the potential difference between two limb electrodes. Leads such as view-18 and view-24 belong to the unipolar lead system, which measure the potential difference between a body electrode and a reference point approximating the heart’s electrical activity. Better in \textbf{bold}.}
\label{tab:panobench performance}
\resizebox{\textwidth}{!}{
\begin{tabular}{c|c|cc|cc|cc|cc}
\toprule
\multirow[c]{3}{*}{\textbf{Input Leads}} & \multirow[c]{3}{*}{\textbf{Method}} 
& \multicolumn{8}{c}{\textbf{Supervised Leads}} \\
 & & \multicolumn{2}{c}{3} & \multicolumn{2}{c}{6} & \multicolumn{2}{c}{9} & \multicolumn{2}{c}{12} \\
 & & PSNR & SSIM & PSNR & SSIM & PSNR & SSIM & PSNR & SSIM \\
\midrule
\multirow{4}{*}{I, II, view-18} 
    & Nef-Net~(Rec) & 35.13 & 0.983 & 34.22 & 0.981 & 34.28 & 0.983 & 34.68 & 0.981 \\
    & \model~(Rec) & \textbf{37.25} & \textbf{0.989} & \textbf{36.22} & \textbf{0.985} & \textbf{36.15} & \textbf{0.985} & \textbf{36.06} & \textbf{0.985} \\
    & Nef-Net~(Syn) & 21.13 & 0.894 & 29.49 & 0.965 & 31.89 & 0.973 & 32.98 & 0.978 \\
    & \model~(Syn) & \textbf{31.63} & \textbf{0.973} & \textbf{33.05} & \textbf{0.973} & \textbf{35.13} & \textbf{0.983} & \textbf{35.57} & \textbf{0.983} \\
\midrule
\multirow{4}{*}{I, II, view-18, 24, 31} 
    & Nef-Net~(Rec) & 35.77 & 0.986 & 34.80 & 0.974 & 34.19 & 0.976 & 35.14 & 0.978 \\
    & \model~(Rec) & \textbf{38.52} & \textbf{0.992} & \textbf{38.18} & \textbf{0.984} & \textbf{36.66} & \textbf{0.987} & \textbf{38.56} & \textbf{0.987} \\
    & Nef-Net~(Syn) & 21.08 & 0.880 & 30.35 & 0.965 & 32.15 & 0.973 & 33.83 & 0.977 \\
    & \model~(Syn) & \textbf{32.01} & \textbf{0.972} & \textbf{33.24} & \textbf{0.972} & \textbf{35.06} & \textbf{0.980} & \textbf{36.11} & \textbf{0.982} \\
\midrule
\multirow{4}{*}{I, II, view-18, 24, 31, 37, 40} 
    & Nef-Net~(Rec) & 35.82 & 0.984 & 32.37 & 0.975 & 33.86 & 0.982 & 35.34 &0.982 \\
    & \model~(Rec) & \textbf{40.85} & \textbf{0.995} & \textbf{37.25} & \textbf{0.988} & \textbf{36.68} & \textbf{0.987} & \textbf{38.48} & \textbf{0.988} \\
    & Nef-Net~(Syn) & 21.78 & 0.917 & 28.06 & 0.964 & 31.94 & 0.976 & 34.47 & 0.980 \\
    & \model~(Syn) & \textbf{32.86} & \textbf{0.973} & \textbf{33.21} & \textbf{0.975} &\textbf{34.57} &\textbf{0.979} &\textbf{36.11} &\textbf{0.985} \\
\bottomrule
\end{tabular}
}
\vskip -0 in
\end{table}

\subsection{Electrocardio Panorama synthesis Evaluation on \dataset}
\label{electrocardio panorama}

Most existing public datasets provide only 12 views, which constrains comprehensive evaluation. To overcome this limitation, we curated \dataset to more thoroughly assess the effectiveness of \model in synthesizing the Electrocardio Panorama. We focus on three central questions: a systematic comparison of \model with Nef-Net in panoramic ECG synthesis, the effect of input lead quantity on reconstruction and synthesis performance, and the effect of supervised lead quantity on reconstruction and synthesis performance.

Given the extensive combinations of input and target views, we align our input configuration with the panorama training settings used for standard 12-lead ECG synthesis. Because the combinations are numerous, we report representative cases in Table~\ref{tab:panobench performance}. Specifically, we select limb lead I and limb lead II as inputs, together with precordial-like signals corresponding to the chest-lead view angles.

\begin{wrapfigure}{r}{0.65\textwidth}  
  \vspace{-1\baselineskip} 
  \centering
  \includegraphics[width=0.62\textwidth]{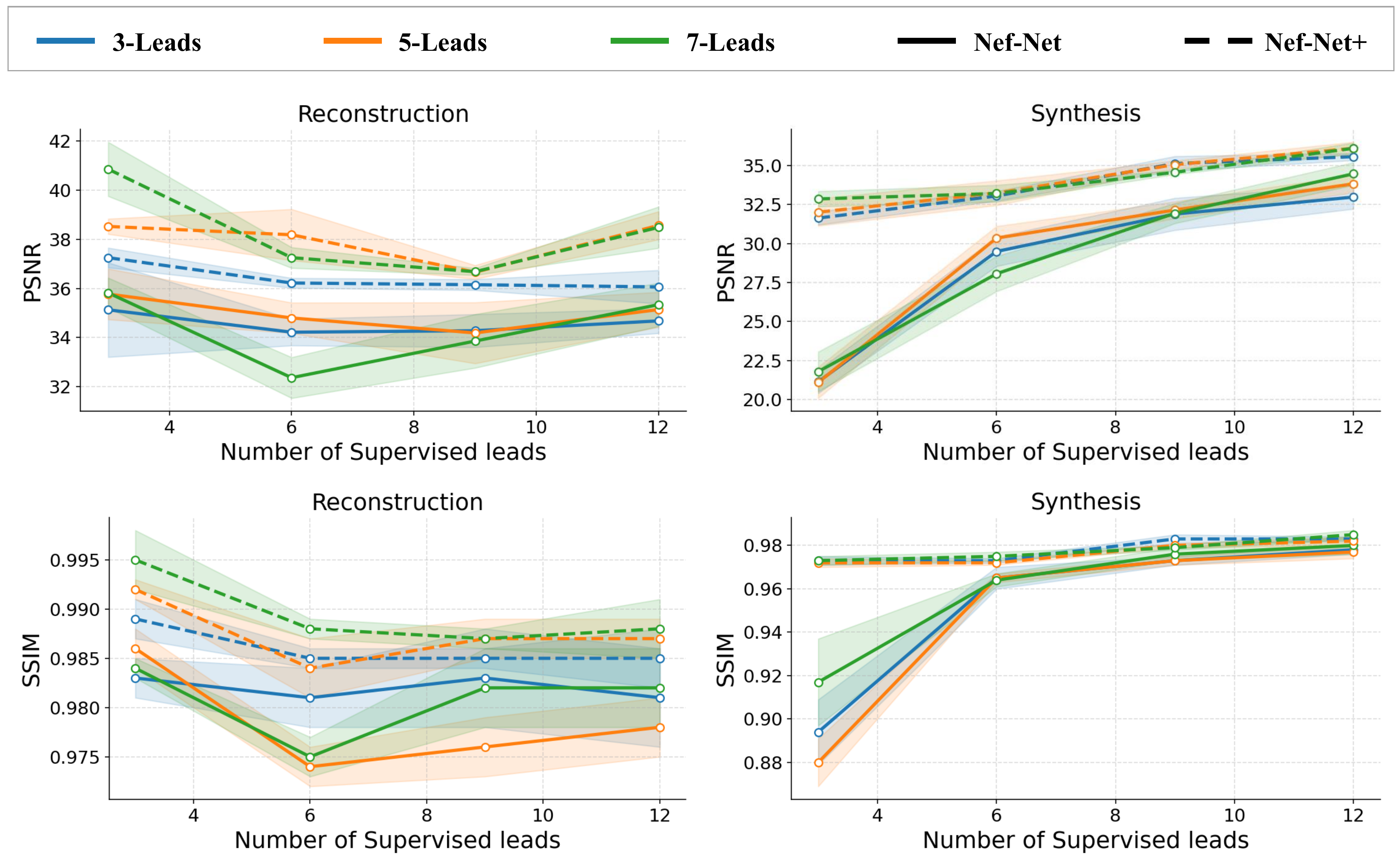}
  \caption{Performance of \model Under Varying Input Conditions and Supervision Leads for Reconstruction and Synthesis Tasks.}
  \label{fig:psnr}
\end{wrapfigure}

Figure~\ref{fig:psnr} compares the synthesis and reconstruction performance of \model and Nef-Net under varying levels of supervision. 
Notably, \model achieves higher accuracy in both tasks.
Across both models, reconstruction performance serves as an upper bound, quantifying the gap between each model's synthesis capability and its inherent representational limits. 
We observe that synthesis performance improves markedly with more supervised leads in both models, while reconstruction remains stable. This narrowing performance gap indicates enhanced generalization, as the model's synthesis capability approaches its representational upper bound.
In contrast, Nef-Net struggles under sparse-lead supervision but gradually improves as more leads are available, eventually approaching \model’s synthesis performance. 
In contrast to the significant role of supervision, the variation in the number of input leads (from 3 to 7) did not yield observable changes in the reconstruction or synthesis performance of either model.

The inferior performance of Nef-Net under sparse supervision stems from its limited use of view-specific information. Its encoder-decoder design compresses signals into a latent cardiac field and uniformly averages recorded features, neglecting the varying physiological relevance of individual views. 
As a result, query-irrelevant signals are mixed into the representation, leading to blurred morphology and degraded fidelity, which becomes more pronounced with fewer supervised leads.
In contrast, \model reformulates ECG synthesis as a direct view-to-view transformation. By leveraging angular relationships, it explicitly models inter-lead dependencies and selectively amplifies geometry-consistent features. This design preserves diagnostically relevant details and sustains high synthesis accuracy, even under sparse-lead supervision.

\subsection{Performances on different diseases}
\label{View synthesis tasks on different diseases}
Multi-lead ECGs capture cardiac electrical activity from spatially distinct perspectives (\eg chest leads V1–V4 reflect the anterior wall). To rigorously assess whether \model can synthesize diagnostically reliable signals across diverse cardiac conditions, we evaluate it on the CPSC2018 dataset, which encompasses nine categories of cardiac disorders, with detailed sample distributions provided in Appendix~\ref{sec:Detailed Information of ECG Dataset}.

Quantitative results in Table~\ref{tab:disease-syn} demonstrate \model's superior reconstruction fidelity across all pathological categories, with an average PSNR improvement of 6.9 dB and consistent SSIM gains. Notably, its performance on Atrial Fibrillation (AF) signals improved by 7.3 dB. Beyond raw fidelity, the consistently larger margins on pathological cases (e.g., AF, I-AVB, PVC, STE) indicate that \model may not only synthesize general signal morphology but also preserves key pathological signatures. The reduced performance gap between normal and abnormal rhythms suggests enhanced disease-adaptive capability, underscoring the potential of \model to generate clinically reliable ECG waveforms under diverse diagnostic scenarios.

\begin{table}[H]
  \centering
  \vskip -0 in
  \caption{View \textit{Synthesis} Performance on CPSC2018 across different diseases. Better in \textbf{bold}.}
\resizebox{\columnwidth}{!}
{
\begin{tabular}{lc|cccccccccc}
\toprule
&& Normal & AF   & I-AVB & LBBB  & RBBB  & PAV   & PVC   & STD   & STE   & AV \\
\midrule
Nef-Net &PSNR & 30.51& 25.12 & 27.35 & 25.24 & 25.19 &27.64 & 28.10 &28.95 & 27.15 & 26.72 \\
\model&PSNR&\textbf{35.41} & \textbf{32.42} & \textbf{33.41}&\textbf{28.35}&\textbf{32.18}&\textbf{33.67}&\textbf{33.14}&\textbf{34.96}&\textbf{33.24}&\textbf{33.62}\\
\midrule
Nef-Net &SSIM  &0.977&0.941&0.961&0.944&0.943&0.959 &0.962&0.965&0.958&0.957\\
\model&SSIM &\textbf{0.989}&\textbf{0.976}&\textbf{0.978}&\textbf{0.955}&\textbf{0.975}&\textbf{0.976}&\textbf{0.982}&\textbf{0.984}&\textbf{0.977}&\textbf{0.985}\\
\bottomrule
\end{tabular}
}
\label{tab:disease-syn}
\vskip -0 in
\end{table}

\subsection{Ablation Study for the Three-Stage Development Framework}
\label{ablation study}
The \pretrain stage establishes a robust baseline by learning from large-scale ECG datasets. The \hospital stage adapts the model to specific ECG acquisition devices, while the \patient stage provides case-level refinements by correcting for electrode placement variability and subject-specific anatomy. 
See Table~\ref{tab:The meaning of training framework}, the results show that incorporating a Device Calibration step yields consistent PSNR gains of 0-1.07 dB and SSIM improvements of 0-0.004 across benchmarks, while adding On-the-Fly Calibration achieves comparable enhancements at 1.75-2.74 on PSNR and 0.001-0.010 on SSIM. This indicates that inter-subject anatomical differences have a more pronounced impact on Electrocardio Panorama quality than inter-device variations. To further investigate this, we analyze the impact of electrode placement offsets on \model's synthesis capability, with detailed results provided in Appendix~\ref{sec:Futher Discussion of this work}.

\begin{figure}[t]
  \centering
  \includegraphics[width=1\linewidth]{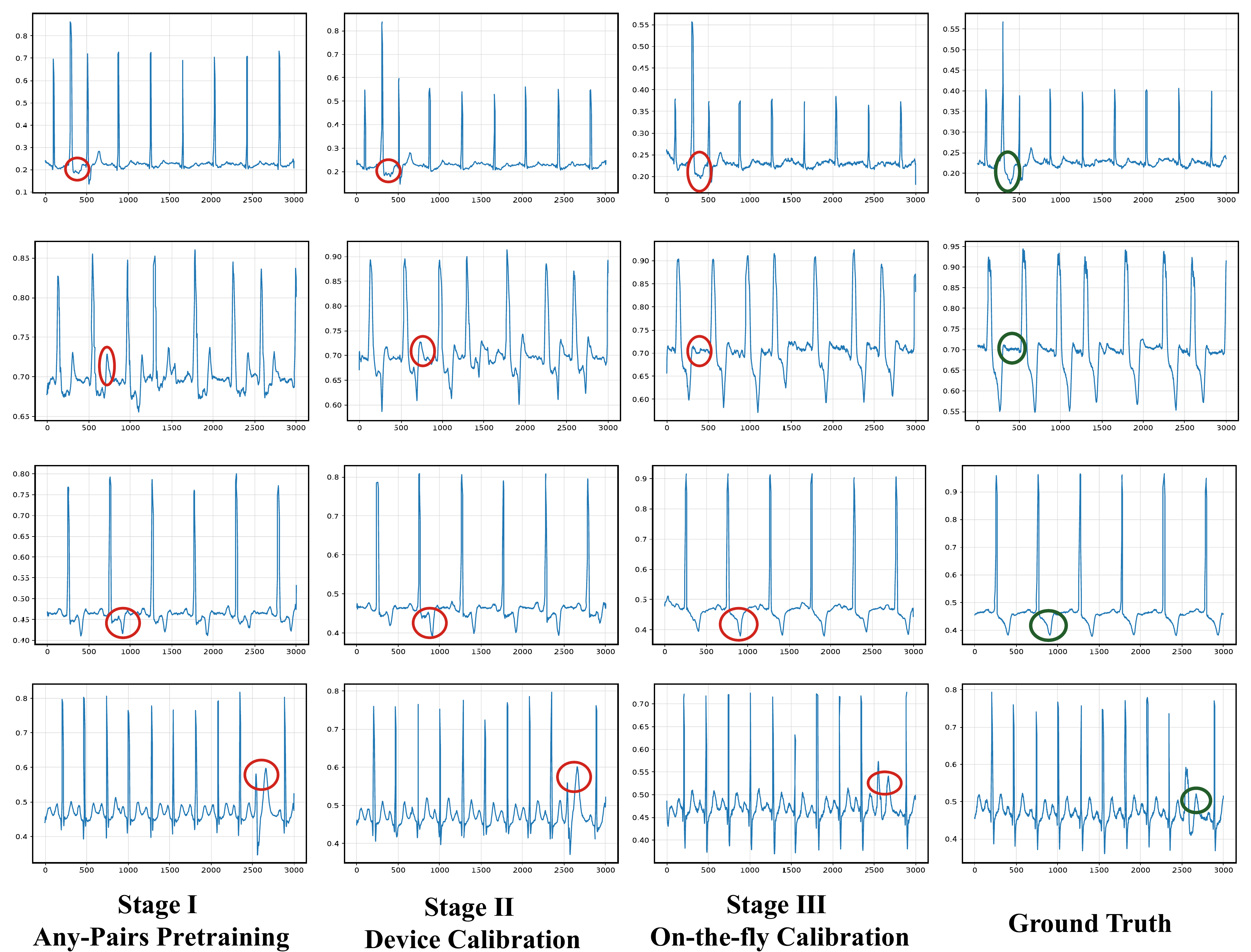}
  \vskip -0.1 in
  \caption{Representative examples from CPSC2018 illustrate progressive synthesis of the V5 view by \model across training stages. Green circles mark clinically relevant diagnostic details in the real signals, whereas red circles highlight their increasingly accurate recovery as our multi-stage training progresses.}
  \label{fig:Signal from different Training stages}
  \vskip -0.2 in
\end{figure}

To further illustrate the benefits of each stage, Figure~\ref{fig:Signal from different Training stages} shows the progressive enhancements achieved by our three-stage framework. 
Importantly, subtle variations in ECG signals are clinically critical, as even minor waveform differences may correspond to arrhythmia events or pathognomonic ST-segment changes. Consequently, even modest improvements in PSNR directly reflect the preservation of clinically salient waveform characteristics that are indispensable for accurate diagnosis and treatment. 
For instance, in Fig.~\ref{fig:Signal from different Training stages} (fourth row), the synthesized signals at Stage I and Stage II exhibit an ST segment higher than the R wave, a morphology that in clinical practice may indicate acute myocardial infarction, thereby potentially misleading diagnosis.
These findings demonstrate the necessity of the complete development-to-deployment framework: \pretrain learns generalizable cardiac spatial priors, \hospital addresses device-level heterogeneity, and \patient adapts to subject-specific anatomical variation. Together, these stages close the gap between controlled training conditions and the demands of real-world clinical deployment.

\begin{table}[t]
\vspace{-0.5em}
  \centering
  \caption{The impact of different stage for view synthesis tasks on the Tianchi, Chinadb, CPSC2018 and PTBXL datasets.}
\resizebox{\columnwidth}{!}{
    \begin{tabular}{c|cc|cc|cc|cc}
\toprule
    \multirow{2}[4]{*}{} & \multicolumn{2}{c|}{ChinaDB} & \multicolumn{2}{c|}{CPSC2018} & \multicolumn{2}{c|}{Tianchi} & \multicolumn{2}{c}{PTBXL} \\
\cmidrule{2-9}          
& PSNR  & SSIM  & PSNR  & SSIM  & PSNR  & SSIM  & PSNR  & SSIM \\
\midrule
    \pretrain &29.83  & 0.972 & 31.01 & 0.975 & 32.71 & 0.975 & 31.15 &0.981 \\
    \hospital & 30.77 & 0.973 & 32.08 & 0.979  & 33.05 & 0.977 &31.15  & 0.981\\
    \patient & 32.57 &0.981  &33.62  & 0.985 & 34.46& 0.976 & 33.41&0.982 \\
\bottomrule
    \end{tabular}
}
  \label{tab:The meaning of training framework}
  \vskip -0 in
\end{table}

\vspace{-0.1 in}
\section{Conclusions}
\vspace{-0 in}
This work advances \textit{Electrocardio Panorama} synthesis from controlled experimental settings to real-world applications. 
Our key methodological contribution is reformulate ECG view synthesis, motivated by heart vector theory and the limitations of Nef-Net’s feature averaging as a direct view-to-view transformation problem. 
Building on this formulation, we introduce a three-stage development pipeline: large-scale pretraining, device-specific calibration, and on-the-fly adaptation to case-specific electrode-induced viewpoint shifts. 
To enable rigorous evaluation, we curate \dataset, the first 48-lead ECG dataset annotated with precise CT-derived spherical coordinates $(\theta,\varphi)$, establishing a comprehensive benchmark for \textit{Electrocardio Panorama} synthesis. Experiments demonstrate that our \model consistently outperforms previous works by a substantial margin.

\newpage


\bibliographystyle{unsrt}  
\bibliography{references} 

\begin{thebibliography}{10}

\bibitem{first-killer}
Thomas Gaziano, K~Srinath Reddy, Fred Paccaud, Sue Horton, and Vivek Chaturvedi.
\newblock Cardiovascular disease.
\newblock {\em Disease Control Priorities in Developing Countries. 2nd edition}, 2006.

\bibitem{2023heart-disease-statistics}
Connie~W Tsao, Aaron~W Aday, Zaid~I Almarzooq, Cheryl~AM Anderson, Pankaj Arora, Christy~L Avery, Carissa~M Baker-Smith, Andrea~Z Beaton, Amelia~K Boehme, Alfred~E Buxton, et~al.
\newblock Heart disease and stroke statistics—2023 update: a report from the american heart association.
\newblock {\em Circulation}, 147(8):e93--e621, 2023.

\bibitem{value-of-12lead-electrocardiogram}
Arnoud~WJ van't Hof, Aylee Liem, Menko-Jan de~Boer, and Felix Zijlstra.
\newblock Clinical value of 12-lead electrocardiogram after successful reperfusion therapy for acute myocardial infarction.
\newblock {\em The Lancet}, 350(9078):615--619, 1997.

\bibitem{kligfield2007recommendationsofecg}
Paul Kligfield, Leonard~S Gettes, James~J Bailey, Rory Childers, Barbara~J Deal, E~William Hancock, Gerard Van~Herpen, Jan~A Kors, Peter Macfarlane, David~M Mirvis, et~al.
\newblock Recommendations for the standardization and interpretation of the electrocardiogram: part i: the electrocardiogram and its technology a scientific statement from the american heart association electrocardiography and arrhythmias committee, council on clinical cardiology; the american college of cardiology foundation; and the heart rhythm society endorsed by the international society for computerized electrocardiology.
\newblock {\em Journal of the American College of Cardiology}, 49(10):1109--1127, 2007.

\bibitem{holter1961new}
Norman~J Holter.
\newblock New method for heart studies: Continuous electrocardiography of active subjects over long periods is now practical.
\newblock {\em Science}, 134(3486):1214--1220, 1961.

\bibitem{van2007posteriormyocardial}
EOF Van~Gorselen, FWA Verheugt, BTJ Meursing, and AJM Oude~Ophuis.
\newblock Posterior myocardial infarction: the dark side of the moon.
\newblock {\em Netherlands Heart Journal}, 15(1):16, 2007.

\bibitem{berne2012brugada}
Paola Berne and Josep Brugada.
\newblock Brugada syndrome 2012.
\newblock {\em Circulation Journal}, 76(7):1563--1571, 2012.

\bibitem{2021-nefnet-electrocardio-panorama}
Jintai Chen, Xiangshang Zheng, Hongyun Yu, Danny~Z Chen, and Jian Wu.
\newblock Electrocardio panorama: synthesizing new ecg views with self-supervision.
\newblock {\em arXiv preprint arXiv:2105.06293}, 2021.

\bibitem{2004-linear}
Stefan~P Nelwan, Jan~A Kors, Simon~H Meij, Jan~H van Bemmel, and Maarten~L Simoons.
\newblock Reconstruction of the 12-lead electrocardiogram from reduced lead sets.
\newblock {\em Journal of electrocardiology}, 37(1):11--18, 2004.

\bibitem{mcculloch1998computationalbiologyoftheheart}
Andrew McCulloch, James Bassingthwaighte, Peter Hunter, and Denis Noble.
\newblock Computational biology of the heart: from structure to function.
\newblock {\em Progress in Biophysics and Molecular Biology}, 69:153, 1998.

\bibitem{2020-rnn}
Andres Hernandez-Matamoros, Hamido Fujita, and Hector Perez-Meana.
\newblock A novel approach to create synthetic biomedical signals using birnn.
\newblock {\em Information Sciences}, 541:218--241, 2020.

\bibitem{2022-lstm}
Ato Kapfo, Sumit Datta, Samarendra Dandapat, and Prabin~Kumar Bora.
\newblock Lstm based synthesis of 12-lead ecg signal from a reduced lead set.
\newblock In {\em 2022 IEEE International Conference on Signal Processing, Informatics, Communication and Energy Systems (SPICES)}, volume~1, pages 296--301. IEEE, 2022.

\bibitem{2022-cnn}
Vishnuvardhan Gundlapalle and Amit Acharyya.
\newblock A novel single lead to 12-lead ecg reconstruction methodology using convolutional neural networks and lstm.
\newblock In {\em 2022 IEEE 13th Latin America Symposium on Circuits and System (LASCAS)}, pages 01--04. IEEE, 2022.

\bibitem{chen2024multi}
Jiarong Chen, Wanqing Wu, Tong Liu, and Shenda Hong.
\newblock Multi-channel masked autoencoder and comprehensive evaluations for reconstructing 12-lead ecg from arbitrary single-lead ecg.
\newblock {\em npj Cardiovascular Health}, 1(1):34, 2024.

\bibitem{seo2022multipleganto12}
Hyo-Chang Seo, Gi-Won Yoon, Segyeong Joo, and Gi-Byoung Nam.
\newblock Multiple electrocardiogram generator with single-lead electrocardiogram.
\newblock {\em Computer Methods and Programs in Biomedicine}, 221:106858, 2022.

\bibitem{golany2019pgans}
Tomer Golany and Kira Radinsky.
\newblock Pgans: Personalized generative adversarial networks for ecg synthesis to improve patient-specific deep ecg classification.
\newblock In {\em Proceedings of the AAAI Conference on Artificial Intelligence}, volume~33, pages 557--564, 2019.

\bibitem{perez2018film}
Ethan Perez, Florian Strub, Harm De~Vries, Vincent Dumoulin, and Aaron Courville.
\newblock Film: Visual reasoning with a general conditioning layer.
\newblock In {\em Proceedings of the AAAI conference on artificial intelligence}, volume~32, 2018.

\bibitem{pipberger1961correlationofdifferentleads}
Hubert~V Pipberger, Stanley~M Bialek, Joseph~K Perloff, and Harold~W Schnaper.
\newblock Correlation of clinical information in the standard 12-lead ecg and in a corrected orthogonal 3-lead ecg.
\newblock {\em American Heart Journal}, 61(1):34--43, 1961.

\bibitem{crossAttention}
Hezheng Lin, Xing Cheng, Xiangyu Wu, and Dong Shen.
\newblock Cat: Cross attention in vision transformer.
\newblock In {\em 2022 IEEE international conference on multimedia and expo (ICME)}, pages 1--6. IEEE, 2022.

\bibitem{hu2018squeeze}
Jie Hu, Li~Shen, and Gang Sun.
\newblock Squeeze-and-excitation networks.
\newblock In {\em Proceedings of the IEEE conference on computer vision and pattern recognition}, pages 7132--7141, 2018.

\bibitem{miyato2018spectral}
Takeru Miyato, Toshiki Kataoka, Masanori Koyama, and Yuichi Yoshida.
\newblock Spectral normalization for generative adversarial networks.
\newblock {\em arXiv preprint arXiv:1802.05957}, 2018.

\bibitem{bailey1990recommendationsstandardizationandspecificationsinautomatedelectrocardiography}
James~J Bailey, Alan~S Berson, Arthur Garson~Jr, Leo~G Horan, Peter~W Macfarlane, David~W Mortara, and Christoph Zywietz.
\newblock Recommendations for standardization and specifications in automated electrocardiography: bandwidth and digital signal processing. a report for health professionals by an ad hoc writing group of the committee on electrocardiography and cardiac electrophysiology of the council on clinical cardiology, american heart association.
\newblock {\em Circulation}, 81(2):730--739, 1990.

\bibitem{schijvenaars2008intraindividualdifference}
Bob~JA Schijvenaars, Gerard van Herpen, and Jan~A Kors.
\newblock Intraindividual variability in electrocardiograms.
\newblock {\em Journal of Electrocardiology}, 41(3):190--196, 2008.

\bibitem{wagner2020ptb}
Patrick Wagner, Nils Strodthoff, Ralf-Dieter Bousseljot, Dieter Kreiseler, Fatima~I Lunze, Wojciech Samek, and Tobias Schaeffter.
\newblock Ptb-xl, a large publicly available electrocardiography dataset.
\newblock {\em Scientific data}, 7(1):1--15, 2020.

\bibitem{cpsc2018}
Feifei Liu, Chengyu Liu, Lina Zhao, Xiangyu Zhang, Xiaoling Wu, Xiaoyan Xu, Yulin Liu, Caiyun Ma, Shoushui Wei, Zhiqiang He, et~al.
\newblock An open access database for evaluating the algorithms of electrocardiogram rhythm and morphology abnormality detection.
\newblock {\em Journal of Medical Imaging and Health Informatics}, 8(7):1368--1373, 2018.

\bibitem{chinadb}
Jianwei Zheng, Jianming Zhang, Sidy Danioko, Hai Yao, Hangyuan Guo, and Cyril Rakovski.
\newblock A 12-lead electrocardiogram database for arrhythmia research covering more than 10,000 patients.
\newblock {\em Scientific data}, 7(1):48, 2020.

\bibitem{E-LSTM}
Jangjay Sohn, Seungman Yang, Joonnyong Lee, Yunseo Ku, and Hee~Chan Kim.
\newblock Reconstruction of 12-lead electrocardiogram from a three-lead patch-type device using a lstm network.
\newblock {\em Sensors}, 20(11):3278, 2020.

\bibitem{KIM}
JA~Kors, G~Van~Herpen, AC~Sittig, and JH~Van~Bemmel.
\newblock Reconstruction of the frank vectorcardiogram from standard electrocardiographic leads: diagnostic comparison of different methods.
\newblock {\em European heart journal}, 11(12):1083--1092, 1990.

\bibitem{anderson1994panoramic}
Stanley~T Anderson, Olle Pahlm, Ronald~H Selvester, James~J Bailey, Alan~S Berson, S~Serge Barold, Peter Clemmensen, Gordon~E Dower, Paul~P Elko, Peter Galen, et~al.
\newblock Panoramic display of the orderly sequenced 12-lead ecg.
\newblock {\em Journal of electrocardiology}, 27(4):347--352, 1994.

\bibitem{case1979sequential}
Robert~B Case, William~A Tansey, and Allen~H Mogtader.
\newblock A sequential angular lead presentation.
\newblock {\em Journal of Electrocardiology}, 12(4):395--401, 1979.

\bibitem{2024-gan-1to12}
Zehui Zhan, Jiarong Chen, Kangming Li, Linfei Huang, Lin Xu, Gui-Bin Bian, Richard Millham, Victor Hugo~C de~Albuquerque, and Wanqing Wu.
\newblock Conditional generative adversarial network driven variable-duration single-lead to 12-lead electrocardiogram reconstruction.
\newblock {\em Biomedical Signal Processing and Control}, 95:106377, 2024.

\bibitem{1950-vector-electrocardiography}
Robert~P Grant.
\newblock Spatial vector electrocardiography: A method for calculating the spatial electrical vectors of the heart from conventional leads.
\newblock {\em Circulation}, 2(5):676--695, 1950.

\bibitem{ITKSNAP}
Paul~A. Yushkevich, Joseph Piven, Heather Cody~Hazlett, Rachel Gimpel~Smith, Sean Ho, James~C. Gee, and Guido Gerig.
\newblock User-guided {3D} active contour segmentation of anatomical structures: Significantly improved efficiency and reliability.
\newblock {\em Neuroimage}, 31(3):1116--1128, 2006.

\end{thebibliography}

\newpage
\newpage
\appendix
\section{Theoretical Foundations for Multi-View ECG Reconstruction} 
\label{sec:Theoretical support}
Electrocardiographic (ECG) signals represent the temporal dynamics of cardiac depolarization and repolarization, recorded as time-series waveforms. A typical cardiac cycle consists of six characteristic deflections: the P wave, PR segment, QRS complex, ST segment, T wave, and TP segment. These deflections reflect distinct physiological processes. For example, the P wave corresponds to atrial depolarization, the QRS complex captures rapid ventricular depolarization, and the T wave reflects ventricular repolarization. Clinically, subtle changes in the amplitude, duration, or morphology of these components can serve as critical biomarkers, such as ST-segment elevation indicating acute myocardial infarction or QRS widening suggesting conduction abnormalities (\eg bundle branch block). This fundamental structure underlies all lead systems and provides the physiological basis for both diagnostic interpretation and computational modeling.

\begin{figure}[htbp]
  \centering
  \includegraphics[width=0.75\linewidth]{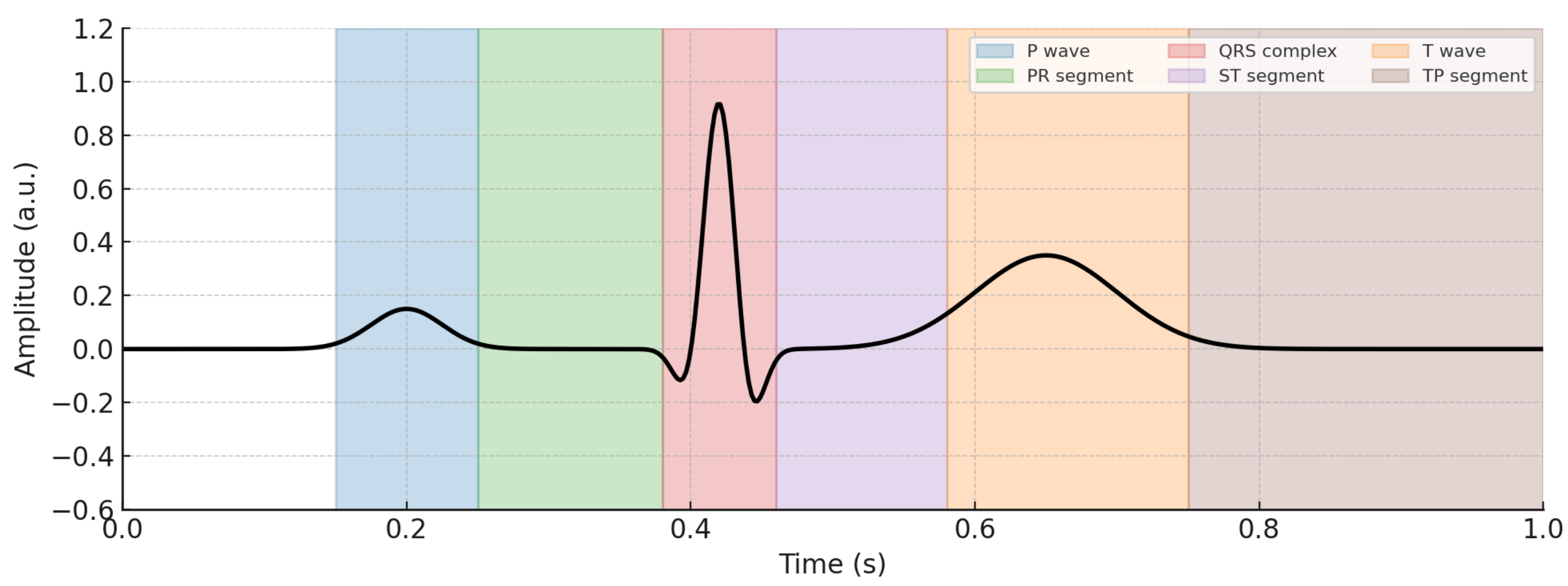}
  \vskip -0 in
  \caption{Standard ECG waveform with six characteristic components highlighted: P wave, PR segment, QRS complex, ST segment, T wave, and TP segment.}
  \label{fig:ECG}
  \vspace{-0.5em}
\end{figure}

In cardiology, ECG signals can be broadly categorized into two lead systems: the bipolar lead system and the unipolar lead system.
The bipolar lead system, including leads \uppercase\expandafter{\romannumeral1}, \uppercase\expandafter{\romannumeral2}, \uppercase\expandafter{\romannumeral3}, as well as augmented leads \( aVR \), \( aVL \), and \( aVF \), measures the potential differences between pairs of electrodes placed on the limbs. These leads offer a rough but global representation of the heart’s electrical activity in the frontal plane. 

In contrast, the unipolar lead system, such as chest leads \( V1, V2, V3, \ldots \) and experimental leads like \texttt{view-1}, \texttt{view-2} in \dataset, record the electrical potential at a specific anatomical location relative to a reference point (often Wilson’s central terminal). The Wilson central terminal, defined as the average of the three limb electrodes, approximates the cardiac reference potential:
\begin{equation}
V_{\mathrm{WCT}} = \frac{V_{\mathrm{RA}} + V_{\mathrm{LA}} + V_{\mathrm{LL}}}{3}
\end{equation}

From a computational perspective, each unipolar chest lead provides a spatially distinct ``viewpoint'' of the heart's electrical activity, analogous to a camera in a multi-view imaging system. This configuration is conceptually similar to a camera network in computer vision, where each camera captures a unique projection of a three-dimensional scene. Here, each lead functions as a biological ``camera'', capturing a unique projection of the heart’s dynamic electrical activity in space and time.
In standard clinical practice, for example, the chest leads V1–V4 are primarily used to assess the anterior wall of the heart, while V5–V6 focus on the lateral wall~\citep{anderson1994panoramic,case1979sequential}. However, the fixed configuration of the 12-lead ECG often fails to capture all diagnostically relevant patterns, as clinicians frequently rely on additional non-standard leads based on individual reading preferences and case-specific requirements~\citep{2024-gan-1to12}. These observations highlight the inherent value of exploring cardiac activity from multiple viewpoints.

\paragraph{Theoretical Foundations of View Transformation for ECG Synthesis.} 
Our model builds upon the principle that each ECG lead represents a projection of the cardiac electrical field from a specific view~\citep{1950-vector-electrocardiography}. 
From an electrophysiological perspective, the cardiac electrical activity can be modeled as a time-varying source whose field is described by a multipolar expansion. As shown in Eq.~\ref{eq:theory}, the extracellular potential $V(x,y,z)$ can be expressed as a series of dipole, quadrupole, and higher-order terms. In clinical practice, higher-order contributions are negligible at the body surface, leading to a far-field dipole approximation in which each ECG lead measures a directional projection of the cardiac dipole vector.
\begin{equation}
\begin{aligned}
    V(x,y,z) &= \frac{1}{4 \pi \sigma}\left(\frac{\sum p_{i} \cdot \hat{r}}{r^{2}}+\frac{\sum\left[(p_{i} \cdot r_{i}^{\prime}) \hat{r}-p_{i}(r_{i}^{\prime} \cdot \hat{r})\right]}{r^{3}}+\cdots\right) 
    \\
    &\approx \frac{1}{4 \pi \sigma}\frac{\sum p_{i} \cdot \hat{r}}{r^{2}} \\
    &\approx \frac{1}{4 \pi \sigma} \frac{p_{x}(x-x_{0})+p_{y}(y-y_{0})+p_{z}(z-z_{0})}{\left[(x-x_{0})^{2}+(y-y_{0})^{2}+(z-z_{0})^{2}\right]^{3 / 2}},
    \label{eq:theory}
\end{aligned}
\end{equation}
where $\sigma$ denotes tissue conductivity, $(p_x, p_y, p_z)$ are the dipole strengths along the cartesian axes, and $(x_0,y_0,z_0)$ denotes the geometric center of cardiac activity, which is not directly measurable. In clinical ECG acquisition, the Wilson central terminal is commonly adopted as a practical approximation to this reference potential. 
This formulation provides the theoretical underpinning for ECG view synthesis: given a limited number of recorded leads, it is possible to infer the cardiac dipole components and reconstruct signals corresponding to arbitrary lead positions $(x,y,z)$ (or equivalently, angular coordinates $(\theta,\varphi)$). Within our model, the relative angles of the input and query leads act as spatial priors, guiding feature aggregation across views. In this context, network parameters implicitly learn conductivity and field distribution properties, enabling physiologically grounded reconstruction of unobserved viewpoints.

From the perspective of cardiac vector theory~\citep{1950-vector-electrocardiography}, clinical ECG acquisition often simplifies this model to a far-field dipole approximation, in which a small subset of independent leads suffices to estimate the global electric field, such that the potential measured at lead $i$ can be expressed as:
\begin{equation}
    V_i(t) \approx \mathbf{p}(t) \cdot \hat{r}_i,
\end{equation}
where $\mathbf{p}(t)$ is the time-varying cardiac dipole vector and $\hat{r}_i$ is the orientation of lead $i$. This formulation highlights that multi-lead ECGs provide discrete directional samples of the same underlying cardiac source.
Synthesizing a novel lead $j$ from recorded leads $\{i\}$ can therefore be viewed as learning a transformation conditioned on angular relationships between $\hat{r}_i$ and $\hat{r}_j$:
\begin{equation}
    V_j(t) \approx \mathcal{T}\big(V_i(t), \hat{r}_i, \hat{r}_j \big),
\end{equation}
where $\mathcal{T}$ is the projection operator implied by the dipole model. Directly estimating this operator from limited data is challenging due to inter-subject anatomical variability and noise. 

Our proposed \model addresses this by embedding angular information $(\theta, \varphi)$ into a learnable representation and coupling it with recorded signal features. This design enables the network to approximate the continuous projection operator $\mathcal{T}$, effectively performing view-to-view transformation without explicitly reconstructing the full cardiac field. In this way, \model operationalizes the dipole-based theoretical foundation into a scalable neural framework capable of synthesizing both standard and novel ECG views.

\section{Dataset Description} 
\label{sec:Detailed Information of ECG Dataset}
We evaluate \model on several widely used public ECG benchmarks. 
\textbf{PTB-XL} contains 21,837 12-lead ECGs, each 10 seconds long and sampled at 500 Hz, covering a broad spectrum of cardiac pathologies. 
\textbf{Tianchi} comprises 31,779 12-lead ECG recordings sampled at 500 Hz. 
\textbf{ChinaDB} includes 10,646 12-lead ECGs (10 seconds, 5,000 samples each) at 500 Hz. 
\textbf{CPSC2018} consists of 6,877 recordings from 11 hospitals, each with 12-lead recordings ranging from 6–60 seconds in length at 500 Hz. 
The CPSC2018 dataset was used to assess \model’s ability to synthesize diagnostically reliable signals under diverse pathological conditions, with Table~\ref{Tab:detailed information of CPSC2018 dataset} summarizing the distribution across nine diagnostic classes. 

\begin{figure}[htbp]
  \centering
  \includegraphics[width=1\linewidth]{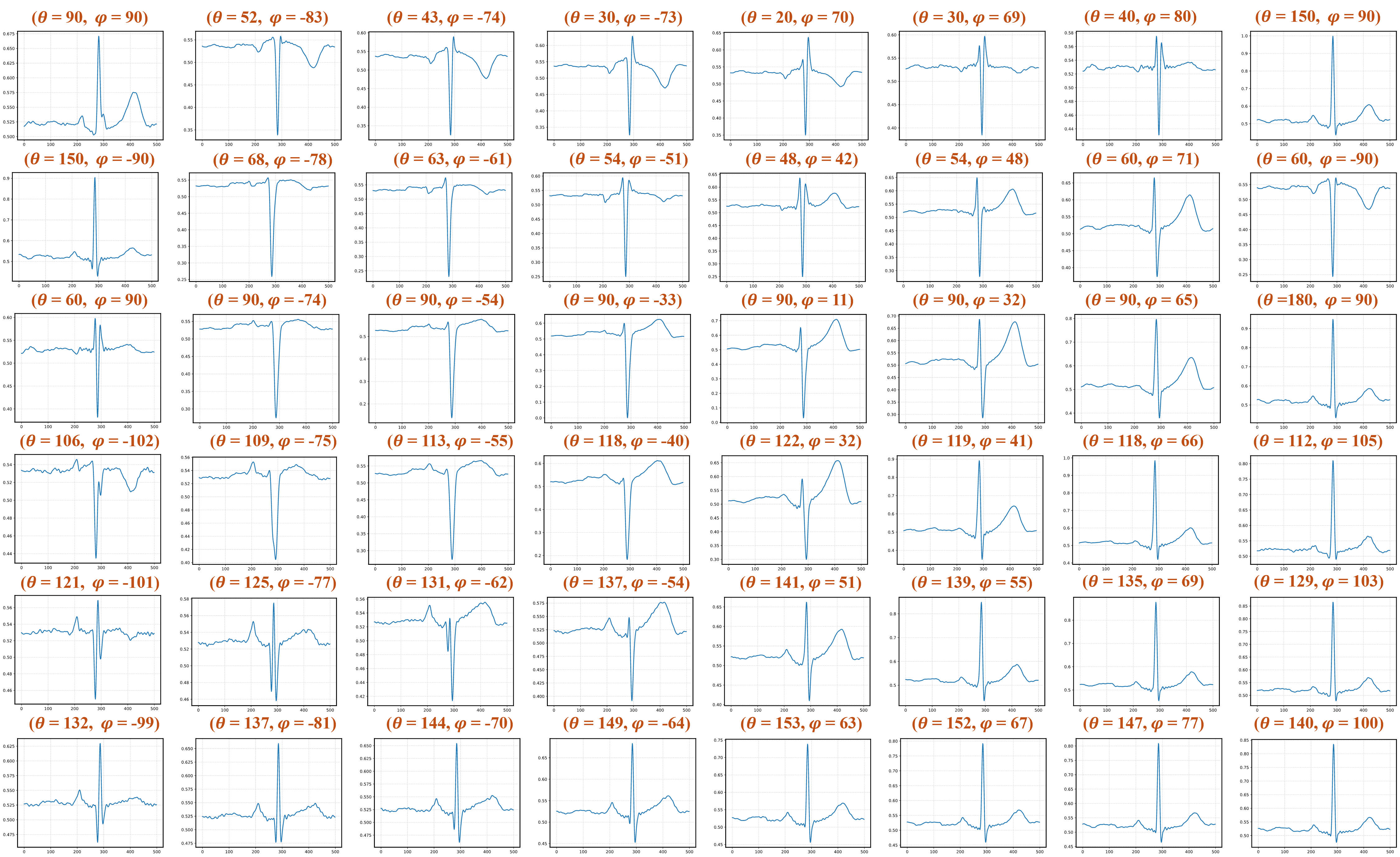}
  \vskip -0 in
  \caption{A representative example from \dataset, illustrating the 48 distinct ECG viewpoints.}
  \label{fig:panobench}
\end{figure}

\begin{figure}[htbp]
  \centering
  \includegraphics[width=1\linewidth]{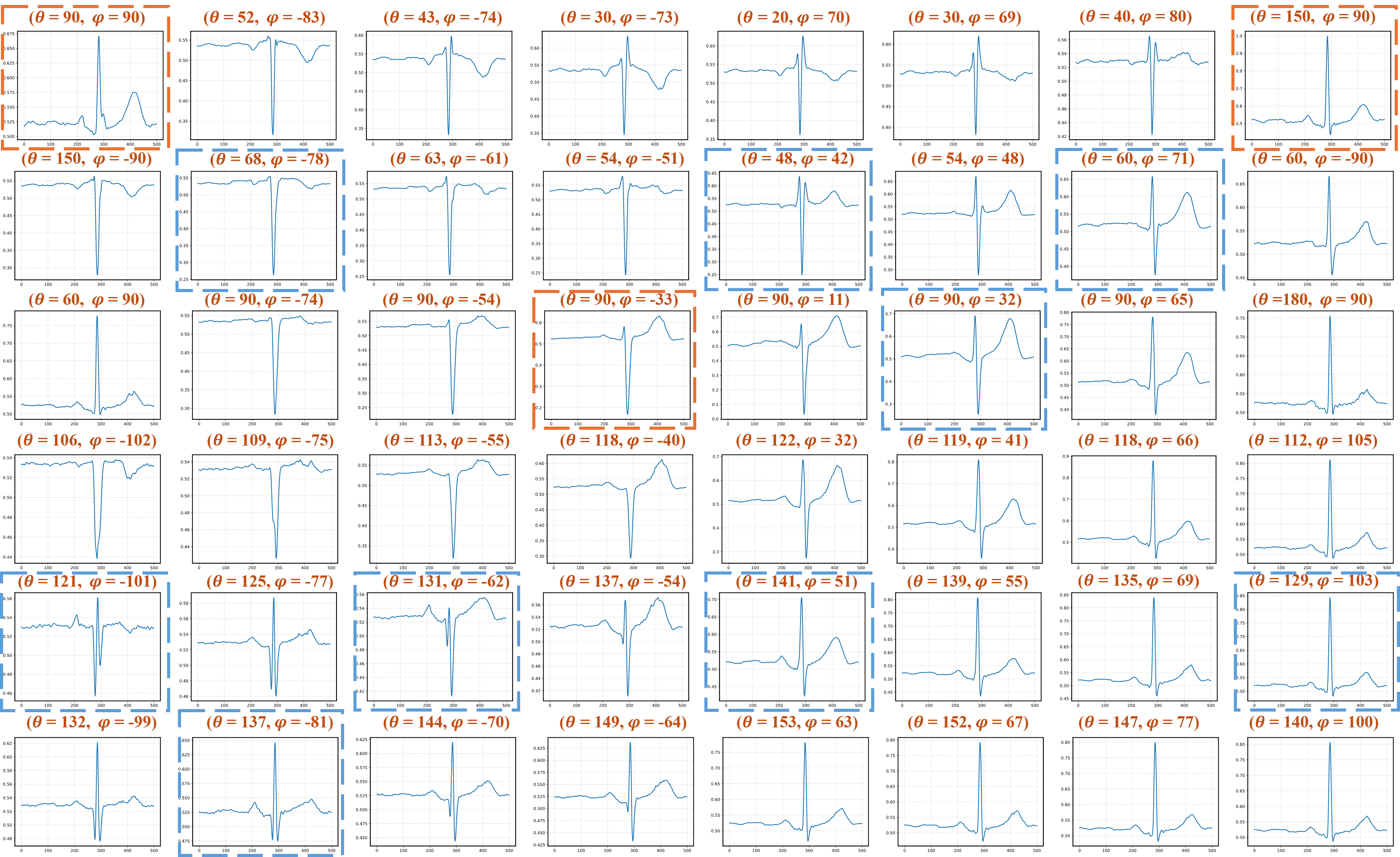}
  \vskip -0 in
  \caption{Representative example from \dataset. Orange dashed boxes denote recorded ECG views, blue dashed boxes mark the views used for supervision during training, and all remaining views are synthesized.}
  \label{fig:panobench_g}
\end{figure}

The \textit{\dataset} dataset is a self-collected ECG resource comprising 5367 recordings (duration 10s), sampled at 250 Hz. Each recording contains 48 leads (6 limb and 42 precordial), collected under resting conditions from subjects aged 18–28. Electrode positions were manually annotated in CT volumes to obtain precise spherical coordinates. The 48-viewpoint signals and their corresponding angular positions are illustrated in Figure~\ref{fig:panobench}; the panorama synthesized for this case is shown in Figure~\ref{fig:panobench_g}. For clarity, orange dashed boxes indicate the recorded ECG views, blue dashed boxes mark the views used for supervision during training, and all remaining views are synthesized. Compared with existing public datasets (\eg PTB-XL, CPSC2018), \dataset provides a substantially denser set of ECG views with explicit angle annotations, offering a unique resource for panoramic ECG synthesis.

A comprehensive comparison of dataset characteristics is provided in Table~\ref{Table:description of dataset}.

\begin{table}[H] 
\vspace{-1.5em}
    \centering
    \setlength{\tabcolsep}{5.5mm}{
    \caption{Detailed Description of the Used ECG Datasets.}
\resizebox{\columnwidth}{!}
{
    \centering
    \begin{tabular}{ccccc}
    \toprule
    Dataset&  Recordings&Sampling Rate&Duration& Lead\\
    \midrule
    PTBXL&21837&500/1000Hz&10s&12\\
    Tianchi&31779&500Hz&10s&12\\
    CPSC2018&6877&500Hz&6-60s&12\\
    ChinaDB&10646&500Hz&10s&12\\
    \dataset&5367&250Hz&10s&48\\
    \bottomrule
    \label{Table:description of dataset}
    \end{tabular}}
}
\end{table}

\begin{table}[H] 
\vspace{-1.5em}
    \centering
    \setlength{\tabcolsep}{5.5mm}{
    \caption{The data description of CPSC2018}
    \resizebox{\columnwidth}{!}
{
    \centering
    \begin{tabular}{ccccccc}
    \toprule
    Class&Training set&\%&Testing set&\%&Total&\%
    \\
    \midrule
    Normal&734&10.6\%&184&2.7\%&918&13.3\%\\
    AF&878&12.8\%&220&3.2\%&1098&16.0\%\\
    I-AVB&563&8.2\%&141&2.1\%&704&10.3\%\\
    LBBB&165&2.4\%&42&0.6\%&207&3.0\%\\
    RBBB&1356&19.7\%&339&4.9\%&1695&24.6\%\\
    PAC&445&6.5\%&111&1.6\%&556&8.1\%\\
    PVC&538&7.8\%&134&1.9\%&672&9.7\%\\
    STD&660&9.6\%&165&2.4\%&825&12.0\%\\
    STE&162&2.4\%&40&0.6\%&202&3.0\%\\
    Total&5501&80.0\%&1376&20.0\%&6877&100.0\%\\
    \bottomrule
    \label{Tab:detailed information of CPSC2018 dataset}
    \end{tabular}}
}
\end{table}

\begin{figure}[htbp]
\vspace{-0em}
  \centering
  \includegraphics[width=1\linewidth]{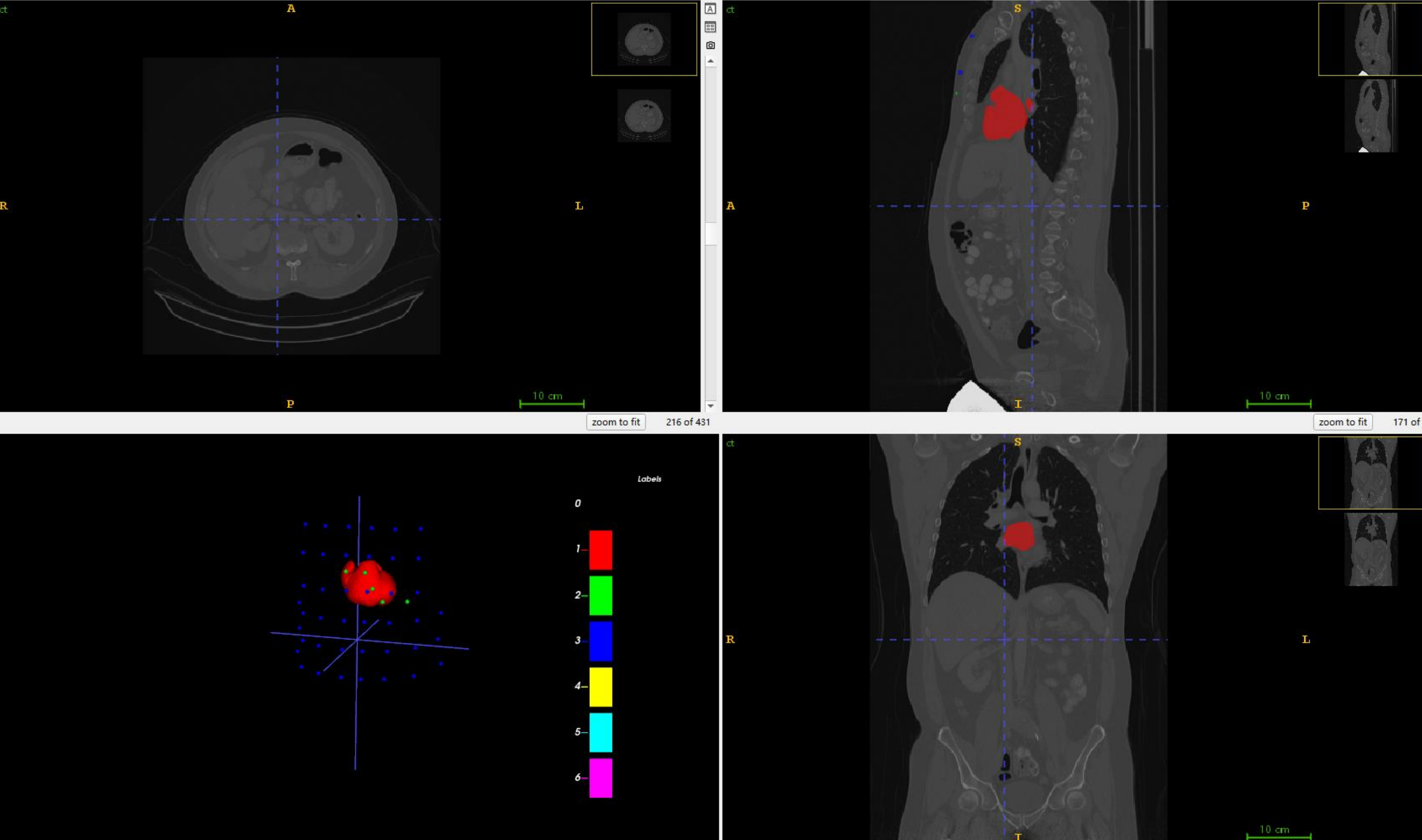}
  \caption{CT Annotation of 48-Lead ECG Electrodes, Six Precordial Leads, and Cardiac Position}
  \label{fig: CT Data Annotation}
\end{figure}

In this work, lead positions were manually annotated using ITK-SNAP software\footnote{www.itksnap.org}~\citep{ITKSNAP}, with the annotation results visualized in Figure~\ref{fig: CT Data Annotation}: the three-dimensional rendering (lower-left panel) highlights the cardiac volume (marked in red as the reference center point), annotated \dataset leads (blue), and standard 12-lead ECG precordial leads (green). A subset of 30 subjects underwent detailed lead angle analysis, and Table~\ref{Angles for panobench} presents the averaged lead angles derived from these annotations. The mean angles of the limb leads are as follows: Lead~\uppercase\expandafter{\romannumeral1}~($90^\circ$, $90^\circ$), Lead~\uppercase\expandafter{\romannumeral2}~($150^\circ$, $90^\circ$), Lead~\uppercase\expandafter{\romannumeral3}~($150^\circ$, $-90^\circ$), aVR~($60^\circ$, $-90^\circ$), aVL~($60^\circ$, $90^\circ$), aVF~($180^\circ$, $90^\circ$).

\begin{table}[htbp]
\vspace{0em}
  \centering
  \caption{Angles for \dataset}
\resizebox{\columnwidth}{!}
{
\renewcommand{\arraystretch}{1.5} 
\begin{tabular}{cc|cc|cc|cc|c|cc|cc|cc|cc}
\toprule
RA & $\theta,\varphi$   &       &  &       &  &       &  & {\multirow{7}[4]{*}{Median}} &       &  &       &  &       & & LA & $\theta,\varphi$  \\

\cmidrule{1-8}\cmidrule{10-17}          & \multicolumn{1}{c|}{} & 4     & 52,-83 & 10    & 43,-74 & 16    & 30,-73 &       & 22    & 20,70 & 28    & 30,69 & 34    & 40,80 &       &  \\

& \multicolumn{1}{c|}{} & 5     & 68,-78 & 11    & 63,-61 & 17    & 54,-51 &       & 23    & 48,42 & 29    & 54,48 & 35    & 60,71 &       &  \\

& \multicolumn{1}{c|}{} & 6     & 90,-74 & 12    & 90,-54 & 18    & 90,-33 &       & 24    & 90,11 & 30    & 90,32 & 36    & 90,65 &       &  \\

1     & 106,-102 & 7     & 109,-75 & 13    & 113,-55 & 19    & 118,-40 &       & 25    & 122,32 & 31    & 119,41 & 37    & 117,66 & 40    & 112,105 \\
    
2     & 121,-101 & 8     & 125,-77 & 14    & 131,-62 & 20    & 137,-54 &       & 26    & 141,51 & 32    & 139,55 & 38    & 135,69 & 41    & 129,103 \\
    
3     & 132,-99 & 9     & 137,-81 & 15    & 144,-70 & 21    & 149,-64 &       & 27    & 153,63 & 33    & 152,67 & 39    & 147,77 & 42    & 140,100 \\
\bottomrule
    \end{tabular}%
}
  \label{Angles for panobench}%
\end{table}

\section{Implementation Details of Different Experiments}
\label{sec:Implementation Details of Different Experiments}
For easy re-implementation, this section documents the experimental configurations and implementation details.

\begin{itemize}
    \item \textit{Number of view transformer layers: 4;}
    \item \textit{0ptimizer: AdamW;}
    \item \textit{Batch size: 32;}
    \item \textit{Weight decay: $10^{-2}$;}
    \item \textit{Learning rate step: [50,100,150];}
    \item \textit{Gamma (a MultiStepLR hype-parameter): 0.5;}
\end{itemize}

\subsection{Any Pairs Pretraining}
200-epoch self-supervised phase employing randomly selected lead combinations (3, 4, or 5 leads per input, excluding synthesized chest leads).
\begin{itemize}
    \item \textit{Training parameters: Exclude all parameters except $\alpha$ (learnable angular correction parameters used only in Stage III);}
    \item \textit{Learning rate: $10^{-3}$;}
    \item \textit{Training Datasets: All datasets;}
    \item \textit{Training epochs: 200}
\end{itemize}

\subsection{Device Calibration}
\begin{itemize}
    \item \textit{Training parameters: Exclude all parameters except $\alpha$;}
    \item \textit{Learning rate: $5*10^{-4}$;}
    \item \textit{Training Datasets: Specific dataset;}
    \item \textit{Training epochs: 200}
\end{itemize}

\subsection{On-the-fly Calibration}
\begin{itemize}
    \item \textit{Training parameters: Exclude all parameters except View Embed block and Reconstruction Head;}
    \item \textit{Learning rate: $5*10^{-5}$;}
    \item \textit{Training Datasets: Per-person;}
    \item \textit{Finetune iterations: 100}
\end{itemize}

\section{Futher Discussion of this work}
\label{sec:Futher Discussion of this work}
\subsection{Impact of \patient on data efficiency for view synthesis}
A data-driven approach is employed to develop the proposed view transformation algorithm, wherein an Any-Pairs pretraining strategy is introduced to enable the model to internalize the “language of ECG signals” while effectively leveraging heterogeneous ECG datasets with diverse lead configurations. Such pretraining is critical for scaling data-driven models to robustly capture cross-lead correlations and improve generalization across varied acquisition settings.

To quantify its impact, we conducted a data-efficiency study by varying the proportion of training data on the CPSC2018 dataset, both with and without pretraining (Table~\ref{tab:abl-datasize}). Results demonstrate that in low-data regimes (1 percent and 5 percent), models trained from scratch exhibit severe performance degradation (e.g., PSNR drop of up to 7.1 dB at 1 percent data), whereas pretraining consistently mitigates this deficit, yielding stable performance even under limited data availability. Beyond 50 percent data, the performance gap narrows, underscoring pretraining’s importance in resource-constrained scenarios and its role in enabling large-scale ECG view synthesis.
\begin{table}[H]
  \centering
  \caption{CPSC2018dataset}
\resizebox{\columnwidth}{!}{
    \begin{tabular}{c|cc|cc|cc|cc|cc}
\toprule
Data volume& \multicolumn{2}{c|}{1\% (44)} & \multicolumn{2}{c|}{5\% (223)} & \multicolumn{2}{c|}{10\% (446)} & \multicolumn{2}{c|}{50\% (2234)} & \multicolumn{2}{c}{100\% (4468)} \\
\midrule
Pretrain   &   NO &  YES     &  NO  &   YES    &  NO   &   YES   &   NO &     YES  & NO & YES\\
\midrule
    PSNR  & 24.68 &31.75& 29.04 & 31.91& 30.05 & 31.89& 31.07 & 32.14& 31.67 & 32.06   \\
    SSIM  & 0.936 &0.978& 0.961 & 0.978& 0.974 &0.979&  0.975&0.979& 0.976 & 0.979       \\
\bottomrule
\end{tabular}%
}
  \label{tab:abl-datasize}
\end{table}

\subsection{Validating \patient under Lead Deviations}
\begin{figure}[htbp]
  \centering
  \includegraphics[width=1\linewidth]{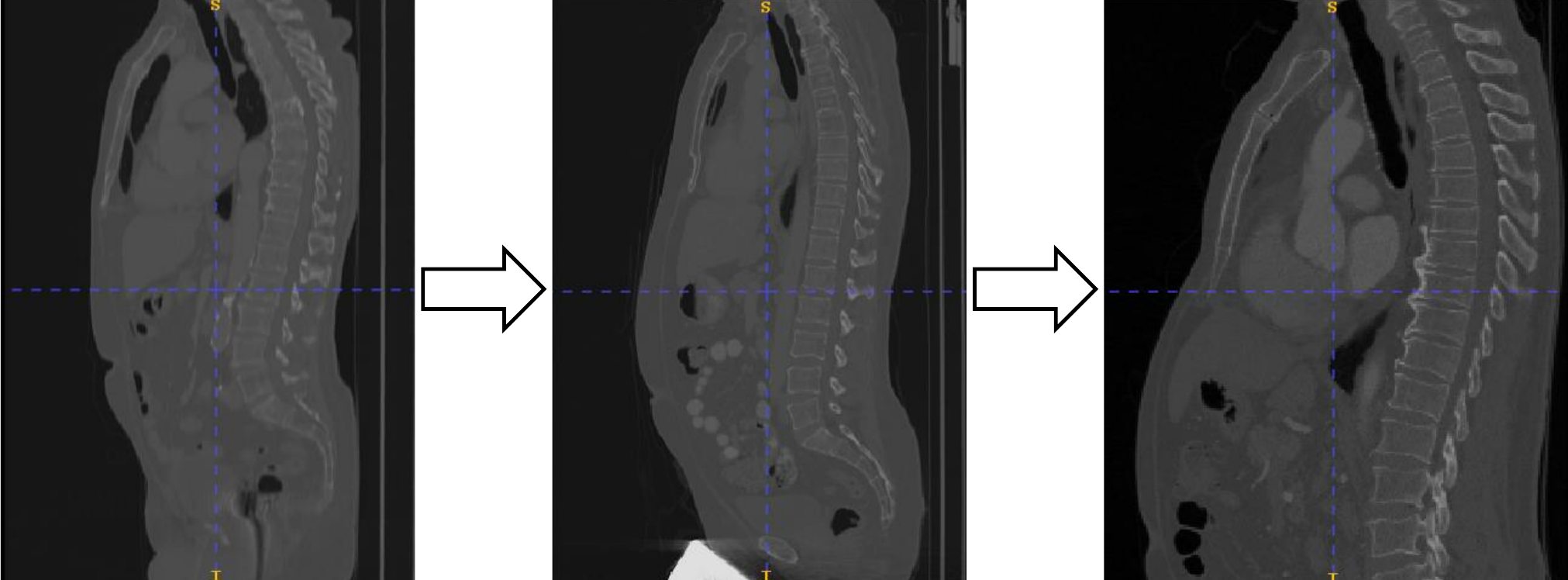}
  \caption{Illustration of inter-individual anatomical variability}
  \label{fig: Anatomical_structure_differences}
\end{figure}

One of the primary objectives of \patient is to perform individual-specific calibration of ECG signals by compensating for deviations in relative lead angles arising from two key sources: (1) variability in electrode placement introduced during manual clinical setup, and (2) inter-individual anatomical differences (\eg variations in heart position and thoracic structure), as illustrated in Fig.~\ref{fig: Anatomical_structure_differences}. 

To evaluate the effectiveness of our learnable angular correction parameter $(d\theta, d\varphi)$, we introduce 10°, 20°, and 30° angular deviation into the CPSC2018 dataset inputs and compare the result before and after the \patient stage. As shown in Table~\ref{tab:angle-correct}, the quality of the result signals drop larger progressively in the absence of correction. Once the model applied $(d\theta, d\varphi)$ and \patient stage, both PSNR and SSIM returned to values close to the unperturbed baseline, demonstrating that our calibration mechanism effectively compensates for electrode misplacement and anatomical variability.

\begin{wraptable}{r}{8cm}
\vspace{-1em}
  \centering
  \caption{The impact of \patient on CPSC2018 (Dataset)}
  \label{tab:angle-correct}
  \begin{tabular}{c|rr|rr}
    \toprule
    \multirow{2}{*}{Deviation} & \multicolumn{2}{c|}{Uncorrected} & \multicolumn{2}{c}{After correction} \\
    \cmidrule{2-5}
    & PSNR & SSIM & PSNR & SSIM \\
    \midrule
    0     & 32.08 & 0.979 & -- & -- \\
    10    & 30.71 & 0.971 & 33.24 & 0.983 \\
    20    & 28.79 & 0.965 & 33.09 & 0.982 \\
    30    & 26.53 & 0.960 & 33.09 & 0.982 \\
    \bottomrule
  \end{tabular}
\end{wraptable}

To achieve individualized model calibration, \model employs an Angle Embed block to encode angular information and a Geometric View Transformer block to perform view transformation. Through adaptive fine-tuning of these modules, \patient aligns recorded ECG views with their anatomically consistent orientations, thereby correcting electrode placement errors and accounting for subject-specific anatomical variability. This calibration step ensures that \model can effectively adapt to individual physiological and acquisition-related differences, ultimately improving its ability to synthesize accurate ECG panoramas across diverse patient populations.

\subsection{Ablation of Model Structure}
Based on the benchmark \model~(denoted as (A) in Table~\ref{tab:abl}), which averages multi-view ECG signals and employs an encoder-decoder architecture to reconstruct the query view via an estimated electric field, we investigate the impact of our proposed components. 

Specifically, (B) incorporates the \textit{Geometric View Transformer} (GeoVT), which explicitly models spatial relationships among ECG views to progressively synthesize the query view, but lacks query-guided encoding. (C) further integrates the \textit{View Encoder} (VEncoder), yielding \model, where query-view angles guide feature extraction: the query embedding functions as an angle-dependent gate that amplifies features aligned with query view while suppressing misaligned ones, thereby enhancing cross-view alignment.

Results show that GeoVT alone improves synthesis PSNR from 26.79~dB to 29.54~dB, demonstrating the effectiveness of explicit geometric modeling. Adding VEncoder yields a substantial further gain (PSNR: 31.19~dB), validating the importance of query-aware feature encoding for precise perspective transformation. From a geometric learning perspective, GeoVT captures inter-view spatial dependencies by progressively aggregating view-consistent features, while VEncoder leverages query-angle embeddings to constrain feature extraction within the correct anatomical frame of reference. Their synergy ensures that synthesized signals remain anatomically consistent and view-coherent, even under significant electrode or anatomical variability. 

Finally, (D) removes the noise perturbation $\epsilon$ and shows a slight performance drop compared to (C), confirming the stabilizing effect of noise injection during training. Overall, these results highlight the complementary contributions of GeoVT, VEncoder, and controlled noise perturbation in improving ECG view synthesis and transformation.

\begin{table}[htbp]
\centering
\caption{Ablation study on CPSC2018 dataset (lead configuration: 3,8,1) evaluated in the \hospital stage.}
\label{tab:abl}
\renewcommand{\arraystretch}{1.2} 
\setlength{\tabcolsep}{0.8em} 
\begin{tabular}{@{}l@{\hspace{1em}}ccc@{\hspace{1.5em}}cc@{\hspace{1.5em}}cc@{}}
\toprule
 & \multicolumn{3}{c}{Components} & \multicolumn{2}{c}{Synthesis} & \multicolumn{2}{c}{Reconstruction} \\
\cmidrule(lr){2-4} \cmidrule(lr){5-6} \cmidrule(l){7-8}
 & $\epsilon$ & GeoVT & VEncoder & PSNR & SSIM & PSNR  & SSIM \\
\midrule
A & \checkmark & -- & -- & 26.79 & 0.958 & 28.47 & 0.960 \\
B & \checkmark & \checkmark & -- & 29.54 & 0.972 & 32.22 & 0.971 \\
C & \checkmark & \checkmark & \checkmark & \textbf{31.19} & \textbf{0.976} & \textbf{35.79} & \textbf{0.981} \\
D & -- & \checkmark & \checkmark & 30.04 & 0.976 & 35.41 & 0.976 \\
\bottomrule
\end{tabular}
\end{table}


\end{document}